\documentclass[journal]{IEEEtran}

\usepackage{amsmath,amssymb,amsfonts}
\usepackage{mathtools}
\usepackage{graphicx}
\usepackage{listings}
\usepackage{hhline}             %nicer double lines in table
\usepackage{multirow}
\usepackage{float}
\usepackage{textcomp}
\usepackage{float}
\usepackage{xcolor}
\usepackage{subcaption}
\DeclareMathOperator{\atantwo}{atan2}

\newcommand{\smashedsqrt}[2][]{%
  \vphantom{#2}%
  \sqrt[#1]{\smash[b]{#2}}%
}

\newcommand*{\figuretitle}[1]{
    {\centering
    \footnotesize #1
    \par}
}

\DeclareCaptionFont{mysize}{\fontsize{8}{9.6}\selectfont}
\begin{document}

\title{A Novel Approach to Vehicle Pose Estimation using Automotive Radar}

%------------------Default author names according to template
% \author{Michael~Shell,~\IEEEmembership{Member,~IEEE,}
%         John~Doe,~\IEEEmembership{Fellow,~OSA,}
%         and~Jane~Doe,~\IEEEmembership{Life~Fellow,~IEEE}% <-this % stops a space
% \thanks{M. Shell was with the Department
% of Electrical and Computer Engineering, Georgia Institute of Technology, Atlanta,
% GA, 30332 USA e-mail: (see http://www.michaelshell.org/contact.html).}% <-this % stops a space
% \thanks{J. Doe and J. Doe are with Anonymous University.}% <-this % stops a space
% \thanks{Manuscript received April 19, 2005; revised August 26, 2015.}}
%------------------------------------------------------------

% \author{Martijn~Heller,
%         Nikita~Petrov,
%         Alexander~Yarovoy}
        
\author{Martijn~Heller\IEEEauthorblockN{\IEEEauthorrefmark{1}, Nikita~Petrov\IEEEauthorrefmark{2} and Alexander~Yarovoy\IEEEauthorrefmark{3}}\\
\IEEEauthorblockA{Microwave Sensing, Signals and Systems (MS3)\\
Delft University of Technology\\
Mekelweg 4, 2628 CD, Delft, the Netherlands\\
E-mails: 
\IEEEauthorrefmark{1}martijn\_heller@hotmail.com, 
\IEEEauthorrefmark{2}N.Petrov@tudelft.nl,
\IEEEauthorrefmark{3}A.Yarovoy@tudelft.nl}
}

%------------------------Default header according to template
% \markboth{Journal of \LaTeX\ Class Files,~Vol.~14, No.~8, August~2015}%
% {Shell \MakeLowercase{\textit{et al.}}: Bare Demo of IEEEtran.cls for IEEE Journals}
%------------------------------------------------------------

% make the title area
\maketitle

% As a general rule, do not put math, special symbols or citations
% in the abstract or keywords.
\begin{abstract}
This paper presents a set of novel scan-matching techniques for vehicle pose estimation using automotive radar measurements. The proposed approach modifies the Normal Distributions Transform (NDT) --- a state-of-the-art scan-matching SLAM technique, widely used in lidar-based localization --- to account for particular aspects of radar environment perception. First, the polar NDT (PNDT) is introduced by solving the NDT  problem in the polar coordinate system, natural for radar measurements. A better agreement between the measurement uncertainties and their representation in the scan-matching algorithm is achieved. Second, the extension of PNDT to take into account the Doppler measurements --- Doppler polar NDT (DPNDT) --- is proposed. %This extension allows for sensor bias estimation in the operational mode.
Third, the SNR of detected targets is added to the optimization procedure to minimize the impact of RCS fluctuation. %After this, the DPNDT is altered to contain an explicit reference to a sensor bias, allowing for its estimation. 
The improvement over the conventional NDT is demonstrated in numerical simulations and real data processing, showing the ability to decrease the localization error by a factor of 3 to 5, depending on the scenario, with a negligible increase in computational complexity. Finally, a DPNDT extension with the capability to compensate angular bias in array beam-forming is presented. Simulation results and real data processing show the possibility to correct it with the accuracy of $0.1^\circ$ almost in real-time. 

% The technique estimates the relative pose of a moving vehicle between different time instances using radar scans. The technique is based on the Normal Distributions Transform (NDT), which was originally designed to perform scan-matching on LIDAR measurements. By adapting the NDT to accommodate radar measurements some of the drawbacks of LIDAR, such as the high cost and poor performance in certain weather conditions, can be overcome. Some of the main disadvantages of using radar as compared to LIDAR, such as lower resolutions, are addressed in the proposed technique. The proposed approach is compared to the NDT using simulations and verified using real data. 
\end{abstract}

% % Note that keywords are not normally used for peerreview papers.
\begin{IEEEkeywords}
pose estimation, localization, sensor calibration, scan matching, Normal Distributions Transform
\end{IEEEkeywords}

\IEEEpeerreviewmaketitle

\section{Introduction}\label{sec:intro}
% The very first letter is a 2 line initial drop letter followed
% by the rest of the first word in caps.
% 
% form to use if the first word consists of a single letter:
% \IEEEPARstart{A}{demo} file is ....
% 
% form to use if you need the single drop letter followed by
% normal text (unknown if ever used by the IEEE):
% \IEEEPARstart{A}{}demo file is ....
% 
% Some journals put the first two words in caps:
% \IEEEPARstart{T}{his demo} file is ....
% 
% Here we have the typical use of a "T" for an initial drop letter
% and "HIS" in caps to complete the first word.

%%--------New intro
\IEEEPARstart{D}{evelopments} in the automotive industry show a trend towards a higher level of automation. Current estimates predict that approximately 2\% of car sales in the UK alone will comprise of vehicles capable of conditional automation. By 2035, it is projected that 40\% of car sales will be Connected and Autonomous Vehicles (CAVs) \cite{forecast}. CAV refers to a vehicle which accommodates autonomy levels of 3 (conditional automation) and above as defined by the SAE International Standard J3016 \cite{SEA}. From this 40\% around 65\% is expected to refer to cars with autonomy levels 4 and 5, meaning high (level 4) to full (level 5) automation will be the standard in these types of vehicles. Examples of high automation are human-activated systems such as automated driving on the freeway with presence of the driver or even automated parking in absence of the driver. Full automation refers to a fully automated driving system without intervention of a human driver \cite{SEA}. These developments pose additional requirements on the on-board sensor systems, as they are expected to accommodate more complex tasks such as object recognition and localization in urban environments \cite{techforauton}. 

The most used approach to perform localization in automotive applications is through the Global Positioning System (GPS). However, especially in urban environments, GPS does not provide high enough accuracy for automated driving application due to multi-path interference and non-line-of-sight reception \cite{SLAMtrends}. A technique known as differential GPS uses ground-based reference stations to improve on accuracy, however the accuracy of such systems is only around 1-3 meters \cite{dGPS}. In the applications of highly and fully automated driving, such low accuracy is unacceptable. Solutions to the localization problem exist that make use of range scanning sensors to estimate the vehicle location within a map of its surroundings. This map can be made a priori \cite{apriorimap} or simultaneous with the estimation of the vehicle location, which is known as the Simultaneous Localization and Mapping (SLAM) problem \cite{SLAMproblem}. The area of SLAM can be divided into two main groups, techniques that perform full SLAM and techniques that perform online SLAM \cite{SLAMtrends}. Full SLAM techniques aim to construct the whole trajectory and map using all measurements and control inputs, whereas online SLAM typically only uses the current sensor inputs. The full SLAM problem is most often solved using the so-called optimization-based approaches. Examples of this are bundle adjustment \cite{bundleadjust1999}, \cite{bundleadjust2005}, and graph SLAM \cite{graph2004}, \cite{graph2007}, \cite{graph2015}. The working principle of optimization-based SLAM solutions is a two-step approach. First, the measurements and control inputs are used to construct constraints after which both the vehicle location and the map are optimized to conform to these constraints. The methods for solving full SLAM are considered to be too complex to perform in real-time \cite{SLAMtrends}. 

Online SLAM methods, on the other hand, are well suited for real-time implementation and are solved using filter-based approaches. These filter-based approaches rely on a prediction step and an update step. The prediction step is based on a dynamic evolution model of the vehicle and the control inputs and produces a predicted estimate of the vehicle and map states. The update step in turn uses the sensor inputs to adjust this estimate. Over the years, a variety of implementations of these so-called Bayesian filters have been proposed, some of which are the Kalman filter \cite{KFSLAM}, the extended Kalman filter \cite{EKFSLAM2000}, \cite{EKFSLAM2001}, the Unscented Kalman filter \cite{UKFSLAM}, and the particle filter \cite{particleSLAM}. Over longer distances, SLAM techniques are known to suffer from drift in their location estimates, this is especially common in the filter-based approaches \cite{SLAMtrends}. A way to compensate for this drift is by using a scan-matching approach. These techniques estimate the pose, i.e. position and heading, of the vehicle relative to a different point in time by maximizing the overlap between the respective range scans. Scan-matching does not require the creation and maintenance of a map of the surroundings \cite{idc}.

Scan-matching techniques use the point clouds that result from 2D or 3D range scans to estimate the relative pose. Range scans at two different time instances are considered, the so-called ``reference scan'' and the ``current scan''. The objective is to find the relative pose of the vehicle between these two scans. The relative pose is estimated by finding the transformation of the current scan point cloud that results in maximum overlap with the reference scan and in turn relating this transformation of the point cloud to the vehicle pose change \cite{psm}. The scan-matching techniques can be divided into two main groups, the feature-based techniques \cite{idc}, \cite{psm}, \cite{icp}, and the distribution-based approaches \cite{ndt}, \cite{3dndt}. The feature-based approaches aim to find the maximum overlap between the scans by minimizing the Euclidean distance between individual features, such as points, lines or surfaces, explicitly. The distribution-based techniques represent the reference scan as a piecewise continuous distribution after which the point of maximum overlap is found by maximizing the probability of the points of the current scan to be at a certain location within this distribution.

Pioneering the distribution-based approaches is the Normal Distributions Transform (NDT) \cite{ndt}. The NDT converts the 2D range scans from the native polar coordinate frame to Cartesian coordinates, after which the reference scan is represented by a combination of bivariate Gaussian distributions related to the distribution of the points within the cells of a grid on the $xy$ plane. The representation of the reference scan on a fixed Cartesian grid comes with drawbacks such as convergence to local minima which have been addressed in proposed techniques \cite{kmeans}, \cite{polarkmeans}. Additional extensions have been proposed to perform scan-matching using three-dimensional scans \cite{3dndt, 3dMagnusson}. 

Due to the high resolution of laser scanners (LiDAR), especially in the angular domain \cite{techforauton, 3dMagnusson, lidarcost}, and due to the popularity of both laser sensors and scan-matching techniques within the robotics community, existing scan-matching techniques have been optimized to accommodate LiDAR technology. Laser scanners, however, come at a higher cost \cite{lidarcost} and perform poorly in bad weather conditions such as heavy rain or fog \cite{lidarweather}. Another line of research is devoted to the application of scan-matching techniques to (stereo) cameras \cite{bertolli2006slam, naikal2009image}, which also become sensitive to weather and lighting conditions when mounted on a car.

These shortcomings of laser and optical sensors can be overcome with a complementary mm-wave radar available in the majority of cars with a high automation level. However, the existing %SLAM and 
scan-matching techniques cannot be applied to radar measurements directly. %Some work has been done on radar SLAM using scan-matching, using multiple previous scans \cite{scanmatchSLAMradar2019, scanmatchSLAMradar2020} or visual features \cite{scanmatchradarvisual2011}. 
The feature-based scan-matching approaches are especially not well suited for radar measurements, due to the fluctuations in radar cross section (RCS) for changing observation angles. 
%Radar measurements are created by transmitting electromagnetic waves and receiving the signals as they are reflected from objects in the surroundings. The amount of signal power that is reflected back is related to the RCS of the object \cite{rcs}. 
The RCS of a target is highly dependent on the observation angle, especially in the case of a complex shape \cite{Principles_of_Modern_Radar}. This in turn results in inconsistency in the reflected power of the same target after movement of the vehicle, which can result in missed detections in one of the scans used for scan matching, causing the point clouds to suffer from ``floating points'' -- the primitive detections which do not have a counterpart in the other scan. Another problem that is encountered when using radar scans is the low angular resolution. In radar systems the angular resolution is determined by the antenna dimensions in terms of the wavelength. For uniform linear arrays (ULA) or MIMO radars with a virtual array, it is determined by the number of  spacial channels. The current generation of automotive radar offers electronic scanning in azimuth realized via 3 Tx $\times$ 4 Rx  MIMO arrays \cite{automotiveInfineon, automotiveNXP}, which can provide angular resolution of an order of $\Delta \theta_{radar} = 5$-$10^\circ$ at the broadside of the radar \cite{angleresradar}, degrading with deviation from the broadside. This is significantly worse than LiDAR measurements, which have angular resolution of an order of $\Delta \theta_{LiDAR} = 1^\circ$ \cite{3dMagnusson}. 
The low angular resolution of radar results in substantial spreading of the target response in the cross-range dimension, depending on target range. The response looks even more complicated after transformation to the Cartesian grid for the conventional NDT, which does not take this effect into account. Moreover, the standard output of the radar (considering standard range-Doppler-angular processing followed by CFAR detection\footnote{FMCW radars have a range estimation bias due to target movements and target migration. We assume that it is properly compensated in the point cloud radar output.}) offers additional information, which can be beneficial for localization, in particular, measurements of the signal-to-noise ratio (SNR) of detected objects and their relative radial velocity with respect to the radar due to Doppler processing.
% the reflected power of each target, providing information about the RCS which can be used to generate a more accurate distribution. Another advantage is the availability of Doppler measurements, these Doppler measurements have a very high resolution, typical resolutions of $\Delta v = 0.05$ m/s if configured for a maximum measurable velocity of $v_{max} = 20$ km/h. 
The access to Doppler measurements firstly provides the possibility to filter out the majority of moving targets --- the ones which have radial velocity --- undesirable in scan matching approaches. Secondly, the access to  Doppler measurements is foreseen to give and improvement in cross-range resolution for stationary targets. In application to forward-looking radars, this processing is referred to as Doppler beam sharpening \cite{dbs, daniel2018application}.
Thirdly, Doppler measurements provide a reference to a target's bearing relative to the motion of the vehicle, which can be exploited to estimate an angular bias in the measurements.
These particular aspects of radar data are however not exploited in the current state-of-the-art  localization techniques.

% Existing scan-matching techniques are not applicable to radar data due to drawbacks associated with radar measurements. Overcoming these drawbacks results in vehicle pose estimation technique that is much more suited for automotive applications than the existing techniques that rely upon LiDAR measurements due to the robustness against bad weather conditions and the lower cost of radar. 

In this paper we propose a set of novel scan matching techniques for vehicle localization using radar data. We recall the scan matching solution of the NDT and address it directly in polar coordinates, native for the sensor. The proposed technique, called polar normal distribution transform (PNDT), allows us to add Doppler measurements into the scan-matching algorithm%, which in turn will present the possibility to perform sensor bias estimation
. We refer to this extension as Doppler polar NDT (DPNDT). Further, we demonstrate the ability to incorporate measurements of target SNR in the localization problem and the benefits of this approach. Moreover, we show that Doppler measurements enforce a particular structure of the data, which gives the potential for radar angular bias estimation and correction %\footnote{Followed by correction of this bias, which lays outside of scope of this paper} 
in the operational mode.

%NDT and develop the solution of scan matching problem directly in native polar coordinates of a sensor by addressing the core problems that arise when using radar data by incorporating the benefits of using radar such as Doppler measurements and received signal power.

% An accurate scan matching technique using radar can even be used to perform sensor bias estimation.

The structure of the paper is as follows: First, the principles of the NDT and its use in finding the maximum overlap are explained in detail in Section \ref{sec:NDT}. Section \ref{sec:PNDT} elaborates on the transformation of the original NDT to accommodate radar measurements. The extension of the DPNDT to perform sensor bias estimation is presented in Section \ref{sec:bias}. Simulation results of the different techniques can be found in Section \ref{sec:sim} and verification using real data is presented in Section \ref{sec:exper}. Finally, in Section \ref{sec:conc}, conclusions are drawn.

\section{The Normal Distributions Transform}\label{sec:NDT}

The objective of scan-matching SLAM techniques is to find the relative pose $\mathbf{p}$ of the vehicle between the two sequential measurements of the environment, referred to as the ``reference scan'' and the ``current scan''. For plane geometry, the relative pose $\mathbf{p} = [t_x, t_y, \phi]^\textrm{T}$ defines the translation of the vehicle between two scans along axes $x$ and $y$ (we consider that axis $x$ aligns with the heading of the vehicle), together with the change of heading $\phi$ between the two scans. 

The Normal Distributions Transform (NDT) is a scan-matching technique, which utilizes a combination of scaled Normal (Gaussian) distributions defined on a regular subdivision of the plane to represent a scan.
This piecewise continuous and twice differentiable description of the map allows for formulation of the objective function as a function of the relative pose $\mathbf{p}$ and for solving this optimization problem using Newton's method. The NDT algorithm is shortly described below; for more details the reader is referred to \cite{ndt}.

\subsection{Scan representation}
% How the distribution is calculated
The conventional implementations of NDT assumes that in each frame the sensor provides a point cloud in 2D Cartesian coordinates corresponding to the detector output. The sensors installed on a car -- lidar, camera and radar -- perceive the environment in polar coordinates; therefore a polar to Cartesian coordinate transform is applied to the data first. 

\begin{figure*}[t]
    \centering
    \begin{subfigure}{0.3\linewidth}
        \figuretitle{Point Cloud}
        \includegraphics[width=\linewidth]{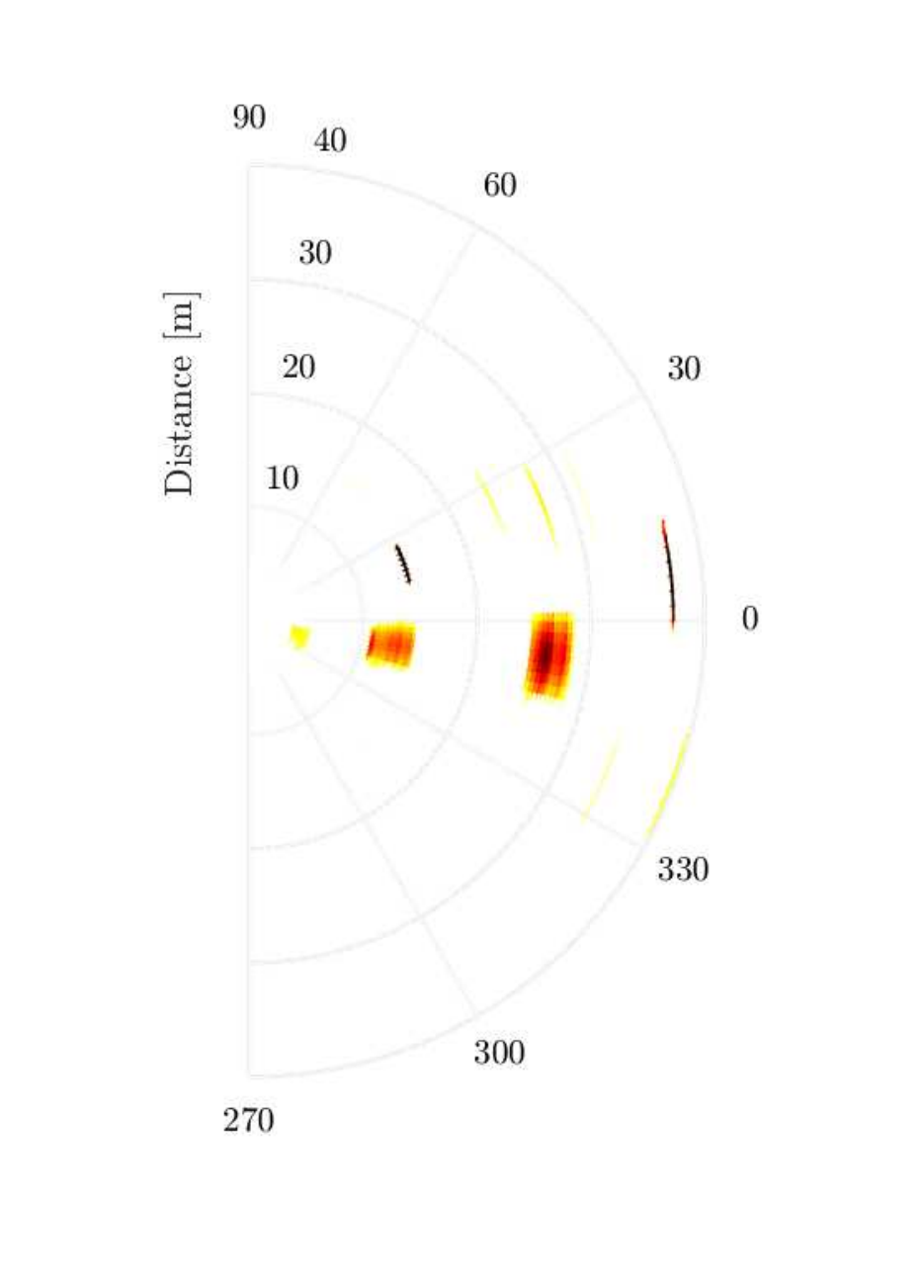}
        \vspace{-3em}
        \caption{}\label{fig:points}
    \end{subfigure}% 
    \hspace{1em}
    \begin{subfigure}{0.28\linewidth}
        \figuretitle{NDT Distribution}\vspace{0.3em}
        \includegraphics[width=\linewidth]{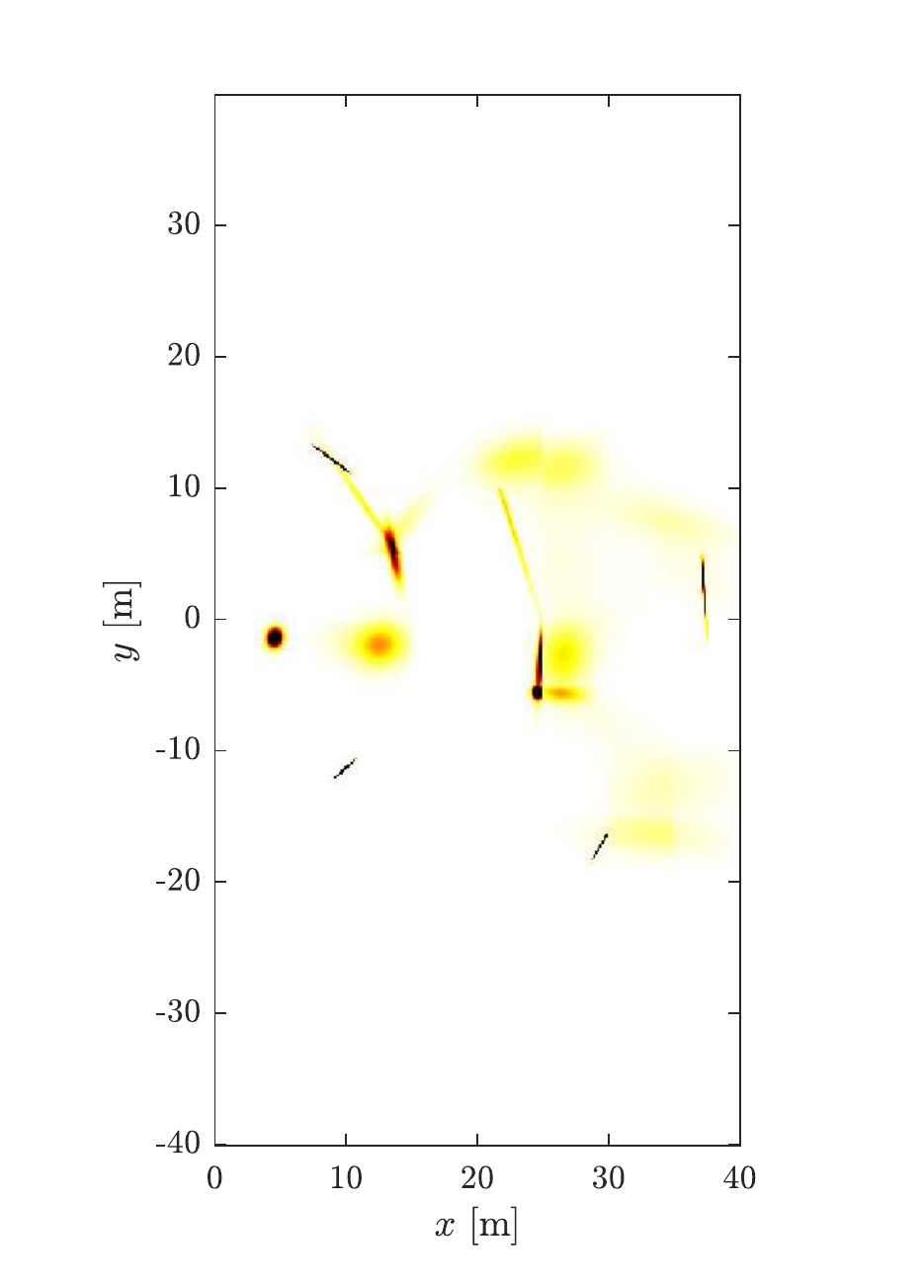}
        \caption{}\label{fig:NDT}
    \end{subfigure}%    
    \hspace{1em}
    \begin{subfigure}{0.3\linewidth}
        \figuretitle{PNDT Distribution}
        \includegraphics[width=\linewidth]{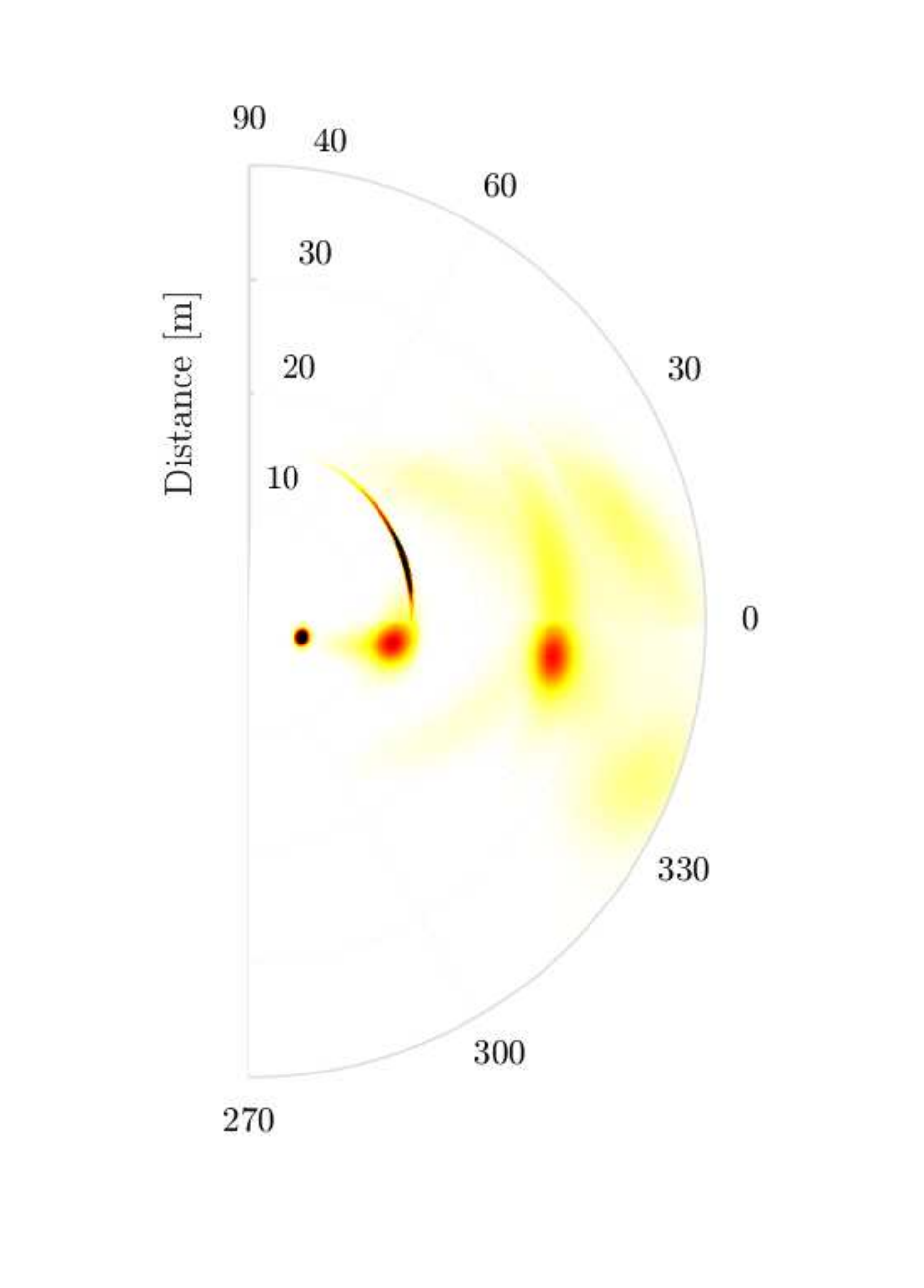}
        \vspace{-3em}
        \caption{}\label{fig:PNDT}
    \end{subfigure}%     
    \caption{Different representations of the point cloud, (a) the original point cloud, (b) the Normal Distributions Transform, (c) the Polar Normal Distributions Transform}\label{fig:scanNDTPNDT}
\end{figure*}

Then, the observed space around the vehicle is divided into equal cells on a regular grid in $x$ and $y$ coordinates.
% In order to construct the piecewise continuous distribution, both the reference and the current scan point clouds, which are measured as range and angle, are converted to the Cartesian coordinate system. 
%Four overlapping grid are constructed on the reference scan. 
For each cell $C_k$ in the grid containing a minimum of three points, the parameters of the bi-dimensional normal distribution -- the mean vector $\mathbf{q}_k \in \mathbb{R}^{2\times1}$ and the covariance matrix $\mathbf{\Sigma}_k \in \mathbb{R}^{2\times2}$ -- are calculated. 
The NDT representation of the reference scan therefore can be seen as a scaled PDF (or likelihood) of the Normal distribution in the grid cell $k$:
\begin{equation}
\label{eq:likilihoodNDT}
\begin{aligned}
L(x, y) \propto \mathcal{N}\left([x, y]^T; \mathbf{q}_k  , \mathbf{\Sigma}_k \right).
\end{aligned}
\end{equation}
To minimize the effect of discetization, four overlapping grids are defined by shifting the original grid by half a cell in $x$ coordinate, $y$ coordinate, and both respectively. The parameters $\mathbf{q}_k$ and $\mathbf{\Sigma}_k$ are estimated for each of them. An example of the NDT representation of an observed scene characterized by the point cloud in Fig. \ref{fig:scanNDTPNDT}, \textit{a} is shown in Fig. \ref{fig:scanNDTPNDT}, \textit{b}.

% This results in a two dimensional mean vector and covariance matrix for each cell inside each of the four grids. The mean vector and covariance matrix will be used as the expected value and covariance in the 2-dimensional Gaussian-based representation of the reference scan. For simplicity, the following parts discuss the steps for only one grid, the steps are repeated for each grid and later summed over the number of grids.

\subsection{The Point Cloud Mapping Equations}\label{subsec:mapping}
% The pose - point cloud transformation relation and explanation about the iterative process

The point cloud of the current scan can be mapped to the coordinates of the reference scan by applying a rotation and translation coordinate transform.
%The mapping of the point cloud of the current scan is done at each iteration using the estimate of the relative pose $\mathbf{p}$ of the previous iteration. 
The position of a point $m$ after applying the pose change  $\mathbf{p} = [t_x, t_y, \phi]^\textrm{T}$ can be found through the mapping equations \cite{ndt}: 
% The dependency of the calculated score in \eqref{eq:probNDT} on $\mathbf{p}$ is thus through these mapping equations.
% 
\begin{equation}\label{eq:mappingeqsNDT}
\begin{aligned}
    x'_m &= x_m \cos(\phi) - y_m \sin(\phi) + t_x; \\
    y'_m &= x_m \sin(\phi) + y_m \cos(\phi) + t_y,
\end{aligned}
\end{equation}
where $[x_m, y_m]$ are the coordinates of point $m$ in the current scan, and $[x'_m, y'_m]$ are the coordinates of point $m$ in the reference scan according to the pose change $\mathbf{p}$.

% For each iteration the, now Cartesian, point cloud of the current scan is transformed using the relative pose $\mathbf{p}$ currently under investigation. $\mathbf{p}$ can be an initialization vector or an estimate from a previous iteration in the optimization process. 

For each mapped point $m$ the corresponding cell in the grid of the reference scan is found and a score is calculated as the value of the scaled Normal probability density function:
\begin{equation}\label{eq:probNDT}
    \textrm{score}(\mathbf{x}'_m) = \exp \left[ - \frac{(\mathbf{x}'_m - \mathbf{q}_k)^\textrm{T} \mathbf{\Sigma}_k^{-1} (\mathbf{x}'_m - \mathbf{q}_k)}{2} \right],
\end{equation}
for $\mathbf{x}'_m = [x'_m, y'_m]^\textrm{T} \in C_k$. 
The objective function of the NDT optimization is then defined as a negative sum of the scores for each point in the current scan:
\begin{equation}\label{eq:objective}
    \min_{\mathbf{p}} \quad - \sum_{m = 1}^M \exp \left[ - \frac{(\mathbf{x}'_m - \mathbf{q}_k)^\textrm{T} \mathbf{\Sigma}_k^{-1} (\mathbf{x}'_m - \mathbf{q}_k)}{2} \right].
\end{equation}
The negative sign is added to formulate a minimization problem
% and is solved using Newton's method. The objective function that is to be minimized is shown in \eqref{eq:objective} \cite{ndt}. This objective function is minimized
to be solved for $\mathbf{p}$, which is involved in \eqref{eq:objective}  through the coordinates mapping \eqref{eq:mappingeqsNDT} of each point $\mathbf{x}'_m$.
% 

% As mentioned earlier, this Gaussian distribution is found using the mean vector and covariance matrix of the reference scan points inside the cell. The score is calculated using \eqref{eq:probNDT} \cite{ndt}. Here, $\mathbf{x}'_m = [x'_m, y'_m]^\textrm{T}$ is a vector containing the mapped coordinates of point $m$ and $\mathbf{q}_m$ and $\mathbf{\Sigma}_m$ are the mean vector and covariance matrix of the cell in which point $m$ resides.

\subsection{The Optimization}
% Details about Newton's method, Hessian, gradient and such
% I think this part might look a little bit too much like the original NDT paper, maybe it is better to refer to that paper for details to avoid plagiarism?
The NDT optimization problem \eqref{eq:objective} is solved using Newton's method. It requires the definition of the gradient and the Hessian of the objective function with respect to $\mathbf{p}$. Data representation via a combination of Normal distributions ensures that both the gradient and the Hessian of the objective function exist at any location within the scan. They are defined from \eqref{eq:probNDT} per data point $m$, respectively, via:
%\textcolor{red}{I did not understand what you want to say in this sentence: The gradient and Hessian of the objective function in terms of the first and second order derivatives of the mapping equations can be found in \eqref{eq:gObj} and \eqref{eq:HObj} \cite{ndt}}. 
%
\begin{equation}\label{eq:gObj}
    \mathbf{g}^i_{m} = \hat{\mathbf{q}}_m^{\textrm{T}} \mathbf{\Sigma}_k^{-1} \frac{\partial \hat{\mathbf{q}}_m}{\partial p_i} \exp \left( - \frac{\hat{\mathbf{q}}_m^{\textrm{T}} \mathbf{\Sigma}_k^{-1} \hat{\mathbf{q}}_m}{2} \right);
\end{equation}
\begin{multline}\label{eq:HObj}
    \mathbf{H}^{i,j}_m = -\exp \left( - \frac{\hat{\mathbf{q}}_m^{\textrm{T}} \mathbf{\Sigma}_k^{-1} \hat{\mathbf{q}}_m}{2} \right) \\ 
    \bigg[ \left( -\hat{\mathbf{q}}_m^{\textrm{T}} \mathbf{\Sigma}_k^{-1} \frac{\partial \hat{\mathbf{q}}_m}{\partial p_i} \right) \left( - \hat{\mathbf{q}}_m^{\textrm{T}} \mathbf{\Sigma}_k^{-1} \frac{\partial \hat{\mathbf{q}}_m}{\partial p_j} \right) + \\
    \left( - \hat{\mathbf{q}}_m^{\textrm{T}} \mathbf{\Sigma}_k^{-1} \frac{\partial^2 \hat{\mathbf{q}}_m}{\partial p_i \partial p_j} \right) + \bigg( - \frac{\partial \hat{\mathbf{q}}_m}{\partial p_i}^{\textrm{T}} \mathbf{\Sigma}_k^{-1} \frac{\partial \hat{\mathbf{q}}_m}{\partial p_j} \bigg) \bigg].
\end{multline}
Here $\hat{\mathbf{q}}_m = \mathbf{x}'_m - \mathbf{q}_k$ with $\mathbf{x}'_m = [x'_m, y'_m]^\textrm{T} \in C_k$, $\mathbf{g}^i_m$ is the $i$-th entry of the gradient vector $\mathbf{g}_m \in \mathbb{R}^{3\times1}$, corresponding to the $i$-th element of $\mathbf{p}$ and $\mathbf{H}^{i,j}_m$ defines the entry on the $i$-th row and $j$-th column of the Hessian $\mathbf{H}_m \in \mathbb{R}^{3\times3}$, corresponding to entries $i$ and $j$ of $\mathbf{p}$. The subscripts $(\cdot)_m$ refer to the calculation of the gradient and Hessian for each data point $m$. The increment step is calculated by averaging gradient and Hessian matrices over the measured point cloud in the current scan and the four overlapping grids to obtain $\mathbf{g}$ and $\mathbf{H}$; see \cite{ndt} for the explicit derivations of \eqref{eq:gObj} and \eqref{eq:HObj}.

% \textcolor{red}{Write explicitly the dimentions of garadinet and Hessian. I think it is $\hat{\mathbf{q}}_m \in \mathbb{R}^{3\times1}$ and $\hat{\mathbf{H}}_m \in \mathbb{R}^{3\times3}$, but the notations if indexies are a bit ambiguous then.}

At each iteration, the current scan point cloud is first transformed according to the mapping equations \eqref{eq:mappingeqsNDT}, then a single Newton step is taken to calculate the increment $\Delta \mathbf{p}$ by solving the following equation: 
\begin{equation}
    \mathbf{H}\Delta \mathbf{p} = -\mathbf{g}
\end{equation}
$\Delta \mathbf{p}$ is added to the previous estimate of $\mathbf{p}$ to obtain a new pose estimate, to be used in the next iteration for mapping the point cloud. This process repeats until convergence.

% \textcolor{red}{Add equation for $\Delta \mathbf{p}$ in the middle of the paragraph above}

\subsection{Discussion}

The NDT provides a powerful tool for scan matching of high-resolution images collected with laser or camera. In radar, the range and angular resolutions are much worse than those of the aforementioned sensors. This leads to spreading of the target responses in these dimensions and significant uncertainty of range and angular measurements of the targets. NDT can be considered as a smoother of these artifacts, but its applicability is limited to the targets representation in the Cartesian grid, while the radar data is naturally represented in polar coordinates. Moreover, conventional automotive radars 
provide the detection cloud supplemented by additional information, such as Doppler velocity and RCS (or signal-to-noise ratio (SNR), which are related to each other by the radar equation), which is not exploited by the standard NDT. 
This motivates formulating and solving the scan-matching problem in polar coordinates.  

%%%%%%%%%%%%%%%%%%%%%%%%%%%%%%%%%%%%%%%%%%%%%%%%%%%%%%%%%%%%%
\section{The Radar Normal Distributions Transform}
\label{sec:PNDT}
%%%%%%%%%%%%%%%%%%%%%%%%%%%%%%%%%%%%%%%%%%%%%%%%%%%%%%%%%%%%

% In order to include the Doppler measurements, the angle of incidence of each target has to be known and mapped. The angle of incidence is available in the original measurements, however, a direct reference to this parameter is lost once the scans are converted to the Cartesian coordinate system. In order to maintain the availability of this angle, the principles of the NDT are applied to the polar coordinate frame, resulting in the Polar Normal Distributions Transform (PNDT).

The necessary changes to allow for the application of the NDT to radar measurements are detailed in this section.

% In this section we apply the principles of the NDT to the sensor (not particularly radar) range and angle measurements. The proposed technique is called the Polar Normal Distributions Transform (PNDT).

\subsection{The Polar NDT (PNDT)}

In this section, we apply the principles of the NDT to the sensor's (not particularly radar) range and angle measurements. The proposed technique is called the Polar Normal Distributions Transform (PNDT). The PNDT performs the methodology of the NDT in polar coordinates:
% In this section we introduce the Polar Normal Distributions Transform (PNDT), which performs all the steps on NDT in polar coordinates: 
% Converting the NDT to polar coordinates requires changes to the algorithm; 
Reference scan representation by means of local Normal distributions is created in polar coordinates and
%distribution is calculated over the polar grid, 
the modified mapping equations relate the range-angular measurements of the current scan to the polar coordinates of the reference scan for the given pose change.  
%used to transform the point cloud of the current scan are changed and in turn 
This leads to new definitions of the Hessian and gradient to be calculated at the optimization step.

\hfill

\subsubsection{The Piecewise Continuous Distribution}
% How the distribution is calculated
The steps in which the distribution is created are the same as for the NDT, the difference being the dimensions of the cells that constitute the grids. That leads to the following PNDT representation of the $k$-th grid cell in the reference scan:
\begin{equation}
\label{eq:likilihoodPNDT}
\begin{aligned}
\tilde{L}(r, \theta) \propto \mathcal{N}\left([r, \theta]^T; \tilde{\mathbf{q}}_k  , \tilde{\mathbf{\Sigma}}_k \right).
\end{aligned}
\end{equation}
Note that despite the similarity with \eqref{eq:likilihoodNDT}, the change of coordinates from Cartesian to polar coordinates completely changes the representation of the map (Fig.  \ref{fig:scanNDTPNDT}, \textit{b}, \textit{c}).
Similarly to the above, four overlapping grids are constructed, each with cells with fixed lengths in the range and angle direction. This causes the surface of the cells to become bigger at larger ranges due to the fixed cell width in degrees. By constructing the grid in this way, the resulting distribution shows a more authentic representation of the received radar measurements: both range and azimuth profiles have a Gaussian shape, which is a reasonable main beam approximation of conventional range-angular processing \cite{Principles_of_Modern_Radar}. 
% in the resulting distribution the power reflected from the targets is represented as a Gaussian distribution around the targets, a Gaussian distribution can be used to approximate the target response in radar measurements %[\textit{does this need a reference?}]
% . Additionally, the spreading of the target response in the cross-range dimension, leading to uncertainty of the angle of incidence of a target for increasing range, is accounted for by the widening of the cells with increasing range. An example of the distribution representation in polar coordinates can be seen in Figure \ref{fig:PNDT}.

\hfill

\subsubsection{The Point Cloud Mapping Equations}
\begin{figure}[t]
    \centering
    \includegraphics[width=0.8\linewidth]{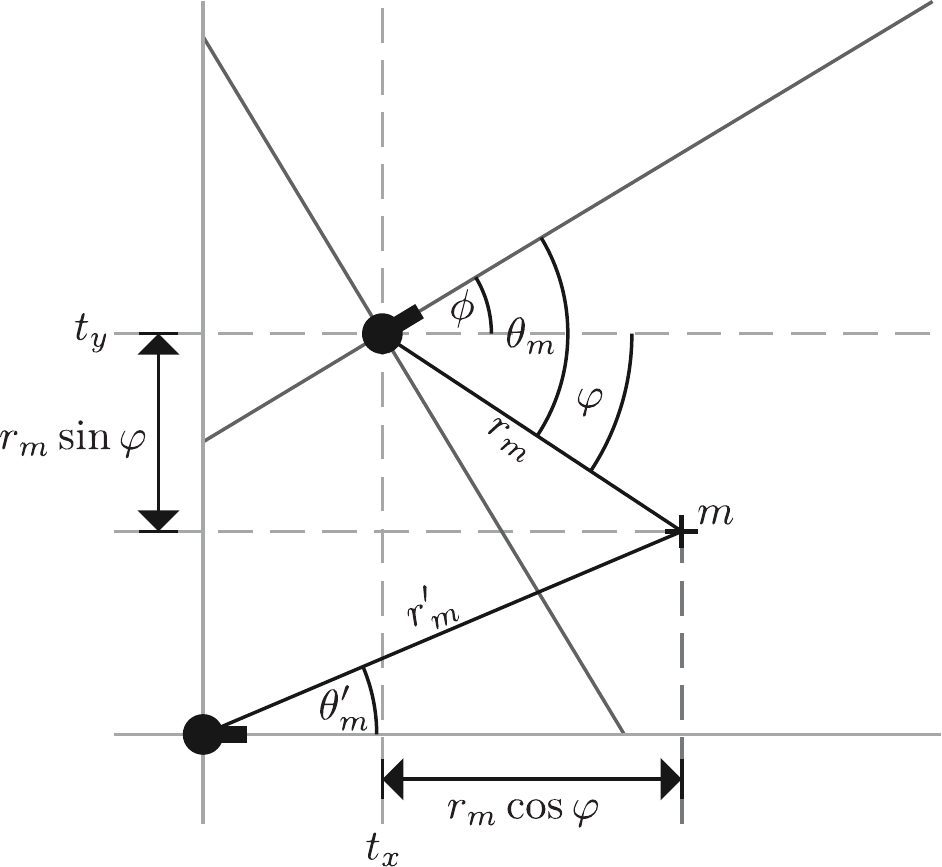}
    \caption{The range-angle measurements of a target $m$ mapped using the relative pose $\mathbf{p} = [t_x, t_y, \phi]^\textrm{T}$. $\varphi$ is calculated as $\phi + \theta_m$. The circles denote the vehicle at two separate positions and the bars indicate the heading of the vehicle, which coincides with the broadside of the radar.}
    \label{fig:mapping}
\end{figure}

% The pose-point cloud transformation relation and explanation about the iterative process
The individual points in the point clouds are now represented by their range $r$ and angle $\theta$. The point cloud of the current scan has to be transformed to the coordinate system of the reference scan given the pose change $\mathbf{p} = [t_x, t_y, \phi]^\textrm{T}$, similar to
%This results in the more involved mapping equations of
\eqref{eq:mappingeqsNDT}. Consider the geometry of the problem depicted in Fig. \ref{fig:mapping}, the mapping equations are given by:
\begin{equation}\label{eq:mappingeqsPNDT}
    \begin{aligned}
        r'_m &= \sqrt{[t_x + r_m \cos (\phi + \theta_m)]^2 + [t_y + r_m \sin (\phi + \theta_m)]^2}; \\
        \theta'_m &= \atantwo[t_y + r_m \sin (\phi + \theta_m), t_x + r_m \cos (\phi + \theta_m)].
    \end{aligned}
\end{equation}
Here, $r_m$ and $\theta_m$ denote the range and angle measurements of point $m$ in the current scan, $r'_m$ and $\theta'_m$ stand for their mapping in the reference scan.
% and $t_x$, $t_y$ and $\phi$ are the values inside $\mathbf{p}$. 
% shows how these mapping equations can be derived, 
The mapping equations of the form \eqref{eq:mappingeqsPNDT} have been used for a point-to-point scan-matching technique in polar coordinates \cite{psm}.

\subsubsection{The Optimization}\label{subsec:optPNDT}
% Details about Newton's method, Hessian, gradient and such
Define the objective function similarly to NDT:
\begin{equation}\label{eq:objective_PNDT}
    \min_{\mathbf{p}} \quad - \sum_{m = 1}^M \exp \left[ - \frac{(\tilde{\mathbf{x}}'_m - \tilde{\mathbf{q}}_k)^\textrm{T} \tilde{\mathbf{\Sigma}}_k^{-1} (\tilde{\mathbf{x}}'_m - \tilde{\mathbf{q}}_k)}{2} \right],
\end{equation}
where $\tilde{\mathbf{x}}'_m = [r'_m, \theta'_m]$, and $\tilde{\mathbf{q}}_k$ and $\tilde{\mathbf{\Sigma}}_k$ are the counterparts of \eqref{eq:objective} for the distribution in polar coordinates.
The optimization is performed with Newton's method, similarly to the NDT.

Since the objective function \eqref{eq:objective_PNDT} has the same form as for the NDT, its gradient and Hessian for each data point $m$ can be expressed similarly to \eqref{eq:gObj}, \eqref{eq:HObj} in terms of the first and second order partial derivatives of the term containing the mapped points ($\hat{\mathbf{q}}_m$ and ${\mathbf{\Sigma}}_k$ being substituted by their counterparts $\hat{\tilde{\mathbf{q}}}_m$ and ${\tilde{\mathbf{\Sigma}}}_k$).
The partial derivatives of the polar measurement equations \eqref{eq:mappingeqsPNDT} are derived in Appendix A.  
% However, the change of the coordinates system and , the mapping equations have changed to the ones in \eqref{eq:mappingeqsPNDT}. The first and second order derivatives of these equations were derived and can be found in the following sections.
These expressions are more involved than those of the conventional NDT \cite{ndt}. This, however, has little impact on the overall computational complexity of the algorithm, as demonstrated below.  
%However, during the verification process it was found that the increase in execution time per iteration of the PNDT compared to the NDT is negligible. Furthermore, the PNDT proves to converge more quickly on average.

% \begin{gather*}
%     \frac{\partial \beta}{\partial \phi} = \frac{r_m^2 + t_x r_m \cos(\phi + \theta_m) + t_y r_m \sin(\phi + \theta_m)}{\left[t_x + r_m\cos(\phi + \theta_m)\right]^2} \\
%     \begin{multlined}
%         \frac{\partial^2 \beta}{\partial \phi^2} = \frac{[t_y r_m \cos(\phi + \theta_m) - t_x r_m \sin(\phi + \theta_m)][t_x + r_m \cos(\phi + \theta_m)]}{[t_x + r_m\cos(\phi + \theta_m)]^3} \\
%         + \frac{2 r_m \sin(\phi + \theta_m) [r_m^2 + t_x r_m \cos(\phi + \theta_m) + t_y r_m \sin(\phi + \theta_m)]}{[t_x + r_m\cos(\phi + \theta_m)]^3}
%     \end{multlined}
% \end{gather*}

%%%%%%%%%%%%%%%%%%%%%%%%%%%%%%%%%%%%%
\subsection{The Doppler Polar NDT (DPNDT)}
%%%%%%%%%%%%%%%%%%%%%%%%%%%%%%%%%%%%%%%%%%%%%%

%------- More focusing on its importance for sensor bias estimation
By addressing the scan matching problem in polar coordinates, a direct reference to the angular measurements becomes available in the mapping equations. This reference can be used to relate the bearing of an object to its measured Doppler velocity. In doing so, the possibility arises to match not only the range and bearing measurements, but also the Doppler measurements. % This approach to the scan matching problem introduces a certain structure between the measured bearing and the Doppler velocity, which can be utilized to estimate a sensor bias.
%-------

% %------- Focussing on increased performance due to higher resolution (which is not supported by the simulation results)
% The performance of scan-matching SLAM significantly depends on the data quality of the sensor data. Thus, relatively low angular resolution of radar and RCS fluctuation of the environment are the essential drawbacks of using radar for localisation in comparison to lidar, especially using range-angular measurement only \cite{scanmatchSLAMradar2019, scanmatchSLAMradar2020}. One way to improve cross-range accuracy of radar measurements consist in joint processing of array beam-forming with Doppler processing, known as Doppler beam sharpening \cite{dbs, daniel2018application}. In this section, we demonstrate how this principle can be incorporated in the NDT localization.
% %-------

% Due to the high resolution and accuracy of Doppler measurements and their relation to the angle of incidence of a target, its incorporation into the scan-matching algorithm can serve as a solution to the low resolution and accuracy of angular measurements in radar.
The relation between the angle of incidence and the Doppler measurements depends on the vehicle's velocity; for a stationary target this relation is given by 
\begin{equation}\label{eq:Dopp}
    v_m = v_\mathrm{car}\cos(\theta_m).
\end{equation}
Here, $v_m$ is the measured Doppler velocity of a stationary target $m$, $\theta_m$ is the target incident angle and $v_\mathrm{car}$ is the velocity of the vehicle in the direction of the broadside of the radar, i.e. in the direction of $t_x$ (Fig. \ref{fig:mapping}).

It is evident from \eqref{eq:Dopp} that in order to incorporate the Doppler measurements and create the Doppler Polar Normal Distributions Transform (DPNDT), an estimate of the velocity of the vehicle should be available. Upon inspection of the geometry in Fig. \ref{fig:mapping} and by assuming a nearly constant velocity, the vehicle velocity can be calculated using the Euclidean distance of $t_x$ and $t_y$, along with knowledge of the frame rate: 
\begin{equation}\label{eq:DoppEuclidean}
    v_m = f \sqrt{t_x^2 + t_y^2} \cos(\theta_m),
\end{equation}
with $f$ the framerate in hertz.
The point cloud in each scan is characterized by their range-angular-Doppler measurements: $\breve{\mathbf{x}}_m = [r_m, \theta_m, v_m]^\textrm{T}$. \\

\subsubsection{The Piecewise Continuous Distribution}
% How the distribution is calculated, now 3-dimensional
The goal is to now optimize the overlap between the range, angle and Doppler measurements of the current scan and the reference scan. For this, the 2-dimensional distribution of the PNDT has to be extended to a 3-dimensional distribution spanning not only the range and angle, but also the Doppler measurements: 
\begin{equation}
\label{eq:likilihoodDPNDT}
\begin{aligned}
\breve{L}(r, \theta, v) \propto \mathcal{N}\left([r, \theta, v]^T; \breve{\mathbf{q}}_k  , \breve{\mathbf{\Sigma}}_k \right).
\end{aligned}
\end{equation}
In the range and angle plane the four overlapping grids are kept and a single additional grid is constructed over the Doppler measurements. The result is four overlapping grids with 3-dimensional cells. For each cell meeting the minimum of three points requirement, again the (now 3-dimensional) mean vector $\breve{\mathbf{q}}_k$ and covariance matrix $\breve{\mathbf{\Sigma}}_k$ are calculated for the reference scan. \\

\subsubsection{The Point Cloud Mapping Equations}
% Added mapping equation of velocity equation
At each iteration, the point cloud of the current scan has to be transformed according to the estimate of the relative pose. In addition to the range and angle measurements \eqref{eq:mappingeqsPNDT}, the Doppler measurements are mapped via:
\begin{equation}\label{eq:mappingeqsDPNDT}
    v'_m = f \sqrt{t_x^2 + t_y^2} \cos(\theta'_m),
\end{equation}
where $\theta'_m$ is the mapped angle according to \eqref{eq:mappingeqsPNDT}.

\subsubsection{The Optimization}
The objective function again takes the shape of \eqref{eq:objective}, \eqref{eq:objective_PNDT}: 
\begin{equation}\label{eq:objective_DPNDT}
    \min_{\mathbf{p}} \quad - \sum_{m = 1}^M \exp \left[ - \frac{(\breve{\mathbf{x}}'_m - \breve{\mathbf{q}}_k)^\textrm{T} \breve{\mathbf{\Sigma}}_k^{-1} (\breve{\mathbf{x}}'_m - \breve{\mathbf{q}}_k)}{2} \right],
\end{equation}
$\breve{\mathbf{x}}'_m$ now denoting the three dimensional measurement vector of point $m$ containing range, bearing and Doppler velocity, and $\breve{\mathbf{q}}_k$ and $\breve{\mathbf{\Sigma}}_k$ the three-dimensional mean and covariance. The optimization steps are the same as previously. For the optimization of the score function, the partial derivatives of the new mapping equations \eqref{eq:mappingeqsDPNDT} involved in the gradient \eqref{eq:gObj} and Hessian \eqref{eq:HObj} are derived in Appendix B.

With these preparations, the Newton optimization can be applied to the range-angular-Doppler point cloud directly.

\subsection{RCS Dependent Scan Matching}
% About the weighted NDT
An additional improvement to the construction of the piecewise continuous distribution is proposed. In addition to the position and Doppler measurements, the power of the received signal can be extracted by means of SNR. %In a noise-limited scenario and known noise power\footnote{The receiver thermal noise can be estimated during the radar silence or well predicted from the operational temperature of the sensor}, radar cross section (RCS) of targets can be directly calculated from the measured SNR \cite{rcs}. 
% In radar measurements, each measured target has an associated reflected signal power. This reflected power is related to the radar cross section (RCS) of the targets. 
Measured SNR can be taken into account at the NDT scan representation and at the optimization step to improve the stability of the algorithm against RCS fluctuations and scintillation of detection points. The received power can be considered as a weight for the measured targets while constructing the mean vector and covariance matrix in each grid cell \eqref{eq:likilihoodNDT}, \eqref{eq:likilihoodPNDT}, \eqref{eq:likilihoodDPNDT}:
\begin{equation}\label{eq:weighted_meancov}
    \begin{gathered}
        \mathbf{q}^i_k = \frac{1}{W_k} \sum^{M_k}_{m = 1} w_m \mathbf{x}^i_m, \\
        \mathbf{\Sigma}^{i,j}_k = \frac{1}{W_k}\sum^{M_k}_{m = 1} w_m (\mathbf{x}^i_m - \mathbf{q}^i_k)(\mathbf{x}^j_m - \mathbf{q}^j_k).
    \end{gathered}
\end{equation}
In these definitions, $i$ and $j$ denote the entries of the vector and matrix, $m$ denotes the measured target whose associated weight $w_m$ is defined as the received power, $M_k$ is the total number of points inside $C_k$ and $W_k =  \sum^{M_k}_{m = 1} w_m$.

Additionally, it can be used to calculate a weighted score for each point, resulting in a weighted sum for the objective function \eqref{eq:objective}, \eqref{eq:objective_PNDT}, \eqref{eq:objective_DPNDT}:
\begin{equation}\label{eq:objectiveSNR}
    \min_{\mathbf{p}} \quad - \sum_{m = 1}^M w_m \exp \left[ - \frac{(\mathbf{x}'_m - \mathbf{q}_k)^\textrm{T} \mathbf{\Sigma}_k^{-1} (\mathbf{x}'_m - \mathbf{q}_k)}{2} \right],
\end{equation}
with the weight factors $w_m$ defined in the same way as in \eqref{eq:weighted_meancov}. This scaling with $w_m$ is a linear operation, it is thus trivial to include it in the Hessian and gradient for the optimization.

\section{Joint Pose and Sensor Bias Estimation}\label{sec:bias}
% About the sensor bias estimation and calibration
An accurate localization with scan matching requires high-quality radar measurements. The performance of the sensors, at the same time, can be affected by many factors. Temperature drift, calibration errors, presence of a water/dirt layer on the bumper or its deformation are examples of factors which can dramatically distort radar performance. Angular measurements, obtained via digital beam-forming, are the most sensitive to these types of distortions, leading to the widening and distortion of the main beam, as well as a bias from its nominal direction \cite{calibration_error}. In this section, we expand the proposed DPNDT solution to the estimation and compensation of radar angular bias.

In order to estimate the angular sensor bias, the properties of Doppler measurements can be utilized. The incidence angle $\theta_{m}$ in \eqref{eq:Dopp} is explicitly defined by the velocity vector of the vehicle and the position of the target $m$. It does not depend on the digital beam-forming and thus has no bias, contrary to the angle $\theta_{m}$ used in \eqref{eq:mappingeqsPNDT}. 
%corresponds to the actual incidence $\theta_{m}$
% Upon close inspection of \eqref{eq:Dopp} it is seen that the measured Doppler velocity of a target is related to its actual, i.e. unbiased, angle of incidence $\theta_{m}$. The angles in \eqref{eq:mappingeqsPNDT} (Fig. \ref{fig:mapping}) are unbiased angles of the targets. 
In this section, we distinguish these two angular measurements by modeling the bias of angular measurements with digital beam-forming by $\theta_{m} = \theta_{m,true} + \epsilon_\theta$, independently on the scanning angle. The modified mapping equations then become:
%
% \begin{equation}\label{eq:mappingeqsPNDTbias}
%     \begin{gathered}
%         r'_m =  \sqrt{[t_x + r_m \cos (\phi + \theta_{m,act})]^2 + [t_y + r_m \sin (\phi + \theta_{m,act})]^2}; \\
%         \theta'_{m,meas} = 
%         \atantwo[t_y + r_m \sin (\phi + \theta_{m,act}), t_x + r_m \cos (\phi + \theta_{m,act})] + \epsilon_\theta; \\
%         v'_m = v_\mathrm{car}\cos(\theta'_{m,act}).
%     \end{gathered}
% \end{equation}
%
\begin{equation}\label{eq:mappingeqsPNDTbias}
    \begin{gathered}
        r'_m = \Big{(} [t_x + r_m \cos (\phi + \theta_{m} - \epsilon_\theta)]^2  \\
        + [t_y + r_m \sin (\phi + \theta_{m} - \epsilon_\theta)]^2 \Big{)}^{1/2}; \\
        \theta'_{m,true} =  \atantwo\Big{[} t_y + r_m \sin (\phi + \theta_{m} - \epsilon_\theta), \\ t_x + r_m \cos (\phi + \theta_{m} - \epsilon_\theta)\Big{]}; \\
        \theta'_{m} = \theta'_{m,true} + \epsilon_\theta; \\
        v'_m = f \sqrt{t_x^2 + t_y^2} \cos(\theta'_{m,true}).
    \end{gathered}
\end{equation}

%Here, $\theta_{m,act} = \theta_{m,meas} - \epsilon_\theta$.
Here, the true, i.e. unbiased, mapped bearing of point $m$ is denoted by $\theta'_{m,true}$. The matching of bearings, however, is based on the biased measurements, which are denoted by $\theta'_{m}$.
By now including the sensor bias $\epsilon_\theta$ in the estimation vector $\mathbf{p}$ and defining the gradient and Hessian for this new parameter, the sensor bias can be found through the standard optimization steps. The gradient and Hessian are closely related to those regarding the heading $\phi$ and are presented in {Appendix C}.

The accuracy of potential bias estimation depends on the available Doppler resolution. To distinguish between any angles $\theta$ and $\theta + \Delta \theta$, the Doppler measurements should be able to distinguish between $v_1 = v_\mathrm{car}\cos\left(\theta\right)$ and $v_2 = v_\mathrm{car}\cos\left(\theta + \Delta \theta\right)$, in other words, the Doppler resolution should satisfy:
\begin{equation}\label{eq:reqDoppRes}
    \begin{gathered}
    % \Delta v = |v_\mathrm{car}\cos\left(\theta\right) - v_\mathrm{car}\cos\left(\theta + \Delta \theta\right)|.
    \Delta v \leq \left|v_\mathrm{car} \left(\cos\left(\theta\right) - \cos\left(\theta + \Delta \theta\right) \right) \right| \\
    = \left|2 v_\mathrm{car} \sin\left(\theta + \frac{\Delta\theta}{2}\right) \sin \left(\frac{\Delta\theta}{2} \right)\right|
    \\ \approx  v_\mathrm{car} \sin\left(|\theta|\right) |\Delta\theta|,
    \end{gathered}
\end{equation}
where we assume that $\Delta\theta$ is small. Substituting the maximum observation angle for $|\theta|$ (typically observation angle is limited to $|\theta| \leq 60^o$) in \eqref{eq:reqDoppRes} defines a simple relation between the achievable accuracy of angle bias correction and radar Doppler resolution. 
It shows that more accurate bias estimation can be archived by increasing the velocity of the vehicle or by improving Doppler resolution. The latter comes at the price of decreased frame rate and may be not feasible in automotive radars with a fixed data processing chain.  Moreover, it explicitly shows that the proposed technique is not applicable in the case of a stationary platform. 

For example, consider an automotive radar with 3 Tx and 4 Rx channels operating at 79 Ghz with the frame rate 40 frames per second (FPS).% and applying time domain multiplexing for beam-forming. 
It provides Doppler measurements with velocity resolution $\Delta v = \lambda / T_\mathrm{frame} \approx 0.15$ m/s, where $\lambda$ is the wavelength at the carrier frequency. Considering maximum observation angle $|\theta| = 60^\circ$ and a velocity of the car $v_\mathrm{car} = 20$ m/s, the achievable accuracy of angular bias estimation is $\Delta \theta \approx \Delta v / \left(v_\mathrm{car} \sin\left(|\theta|\right) \right) \approx 0.5^\circ$.

% In real data applications, a higher resolution can be achieved by concatenating measurements of consecutive frames in the slow time. In doing so, a higher frame rate is sacrificed for higher Doppler resolution. This, however, should not pose a problem as the estimation of a sensor bias need not be performed continuously. The pose estimation can be performed at the higher frame rate after compensation of the sensor bias, which is estimated only occasionally.

\section{Simulations}\label{sec:sim}
% Set-up and results of the simulations
For the verification of the proposed techniques and the comparison to the conventional NDT, a ground truth of the vehicle location is required. In Section \ref{sec:exper} the procedures and results of an experiment using an automotive radar in a dense environment are presented. In this experiment, GPS data is used as a reference. However, as mentioned in Section \ref{sec:intro}, the accuracy of GPS is not sufficient to function as a ground truth, even more so in a dense environment. To test the algorithms with reference to a ground truth, realistic simulations of radar data were undertaken and used for performance assessment of the proposed techniques.
% The setup of the simulations and the performance of the proposed techniques will be discussed in this section. 

\subsection{Simulating Radar Measurements}
\begin{figure}[t]
    \centering
    \includegraphics[width=0.8\linewidth]{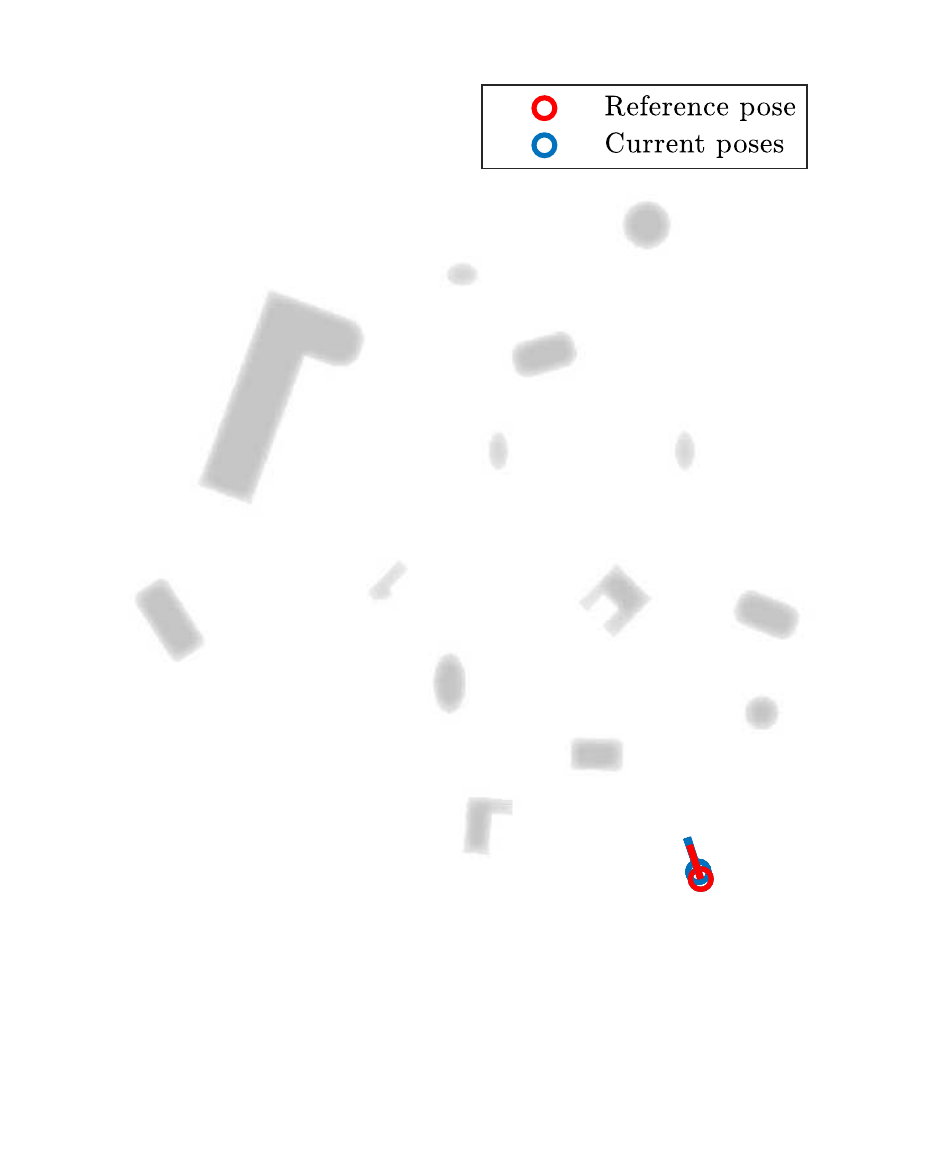}
    \vspace{-5em}
    \caption{The generated map along with the starting pose (red) and the randomly generated relative poses (blue).}
    \label{fig:map}
\end{figure}

A gray-scale map is created of the surroundings of a vehicle. Different shapes are used to simulate different types of objects, each object being characterized by a certain RCS which is represented by its gray-scale value. The resulting map is shown in Fig. \ref{fig:map}. Within this map, a starting position is chosen which will serve as the reference pose. 

\begin{figure*}[t]
    \centering
    \begin{subfigure}{0.33\linewidth}
        \figuretitle{Cell size 2 m}
        \includegraphics[width=\linewidth]{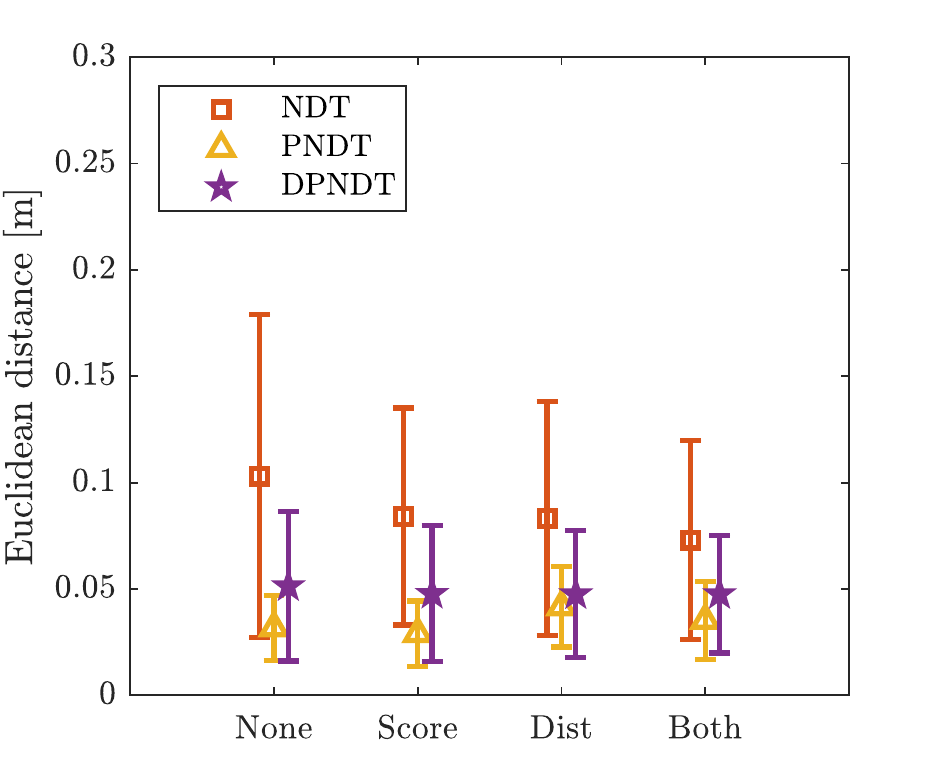}
        \vspace{-2em}
        \caption{}\label{fig:posesDist2}
        \par \medskip
        \includegraphics[width=\linewidth]{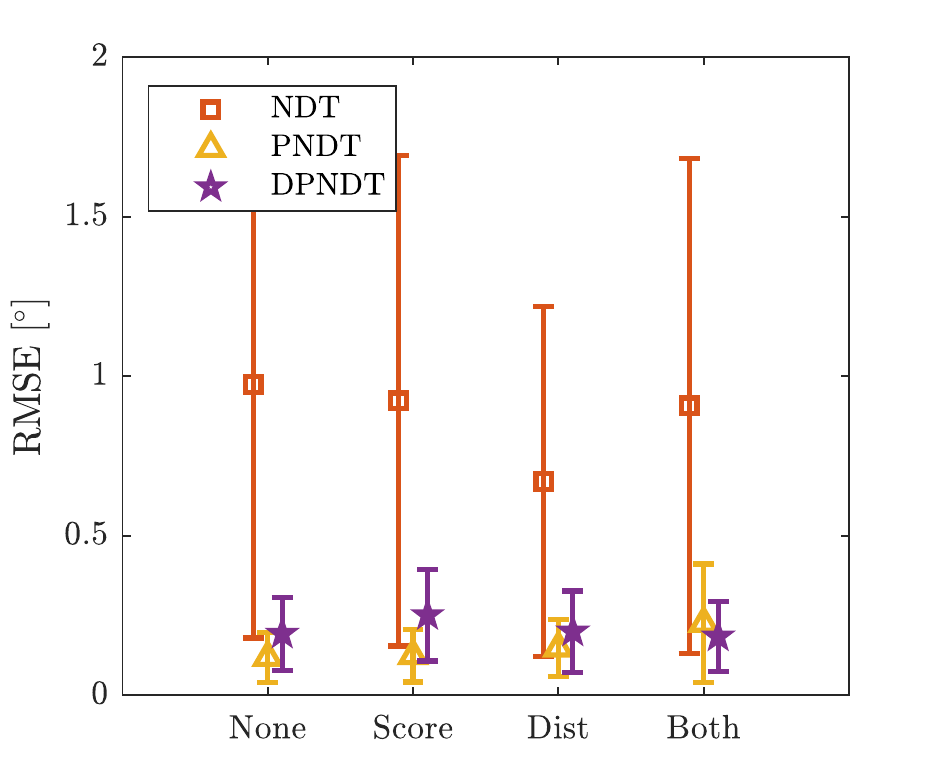}
        \vspace{-2em}
        \caption{}\label{fig:posesHead2}
    \end{subfigure}%         
    \begin{subfigure}{0.33\linewidth}
        \figuretitle{Cell size 3 m}
        \includegraphics[width=\linewidth]{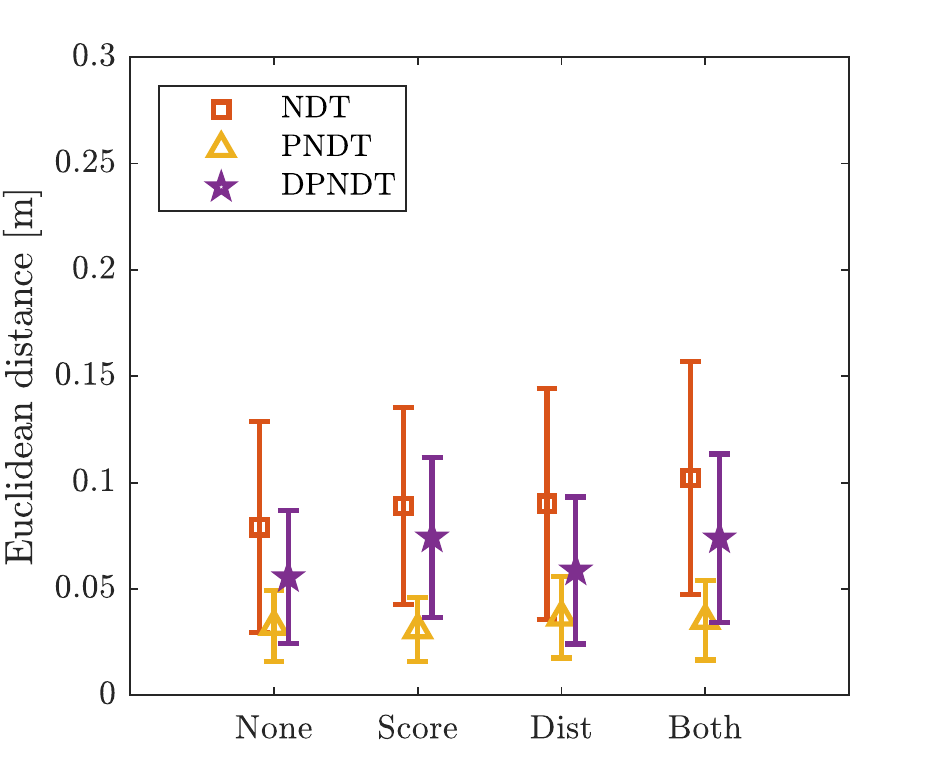}
        \vspace{-2em}
        \caption{}\label{fig:posesDist3}
        \par \medskip
        \includegraphics[width=\linewidth]{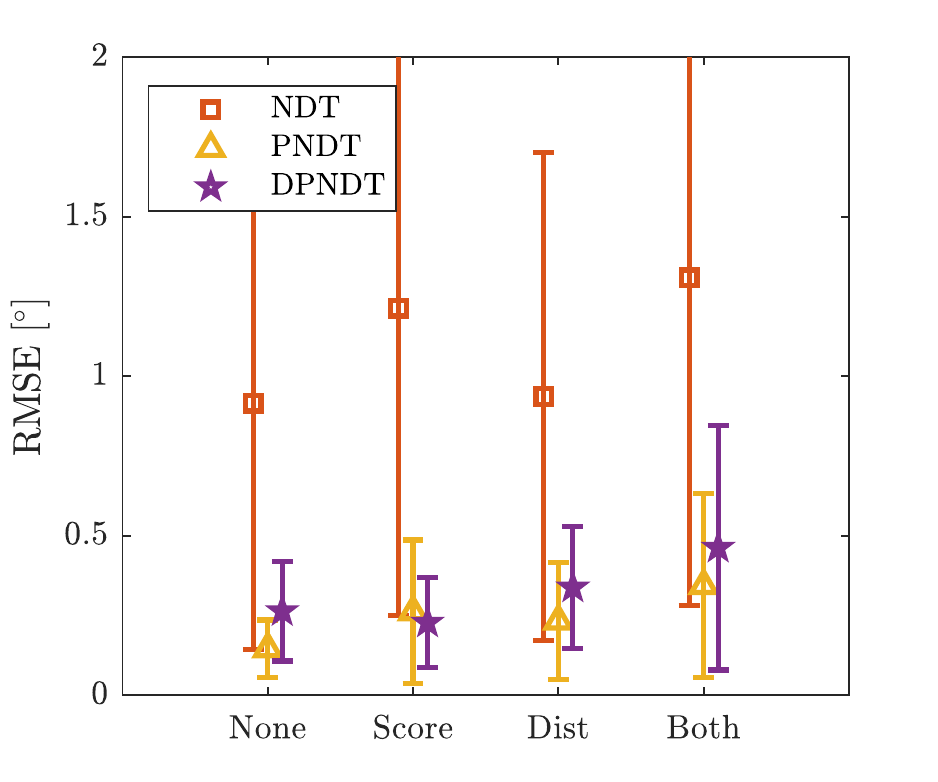}
        \vspace{-2em}
        \caption{}\label{fig:posesHead3}
    \end{subfigure}%         
    \begin{subfigure}{0.33\linewidth}
        \figuretitle{Cell size 5 m}
        \includegraphics[width=\linewidth]{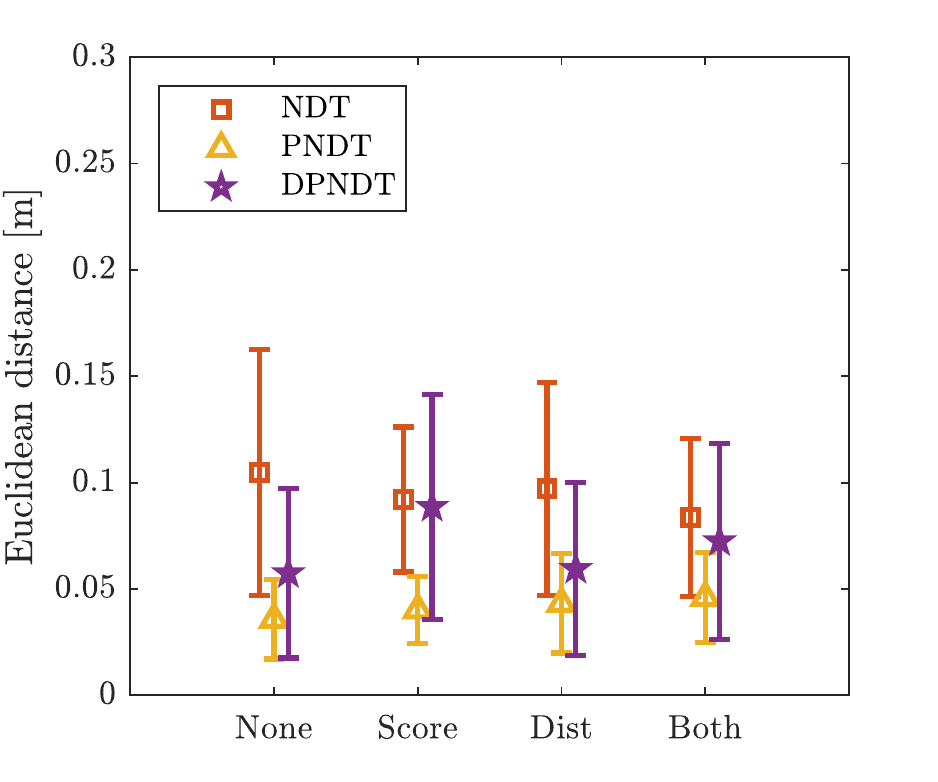}
        \vspace{-2em}
        \caption{}\label{fig:posesDist5}
        \par \medskip
        \includegraphics[width=\linewidth]{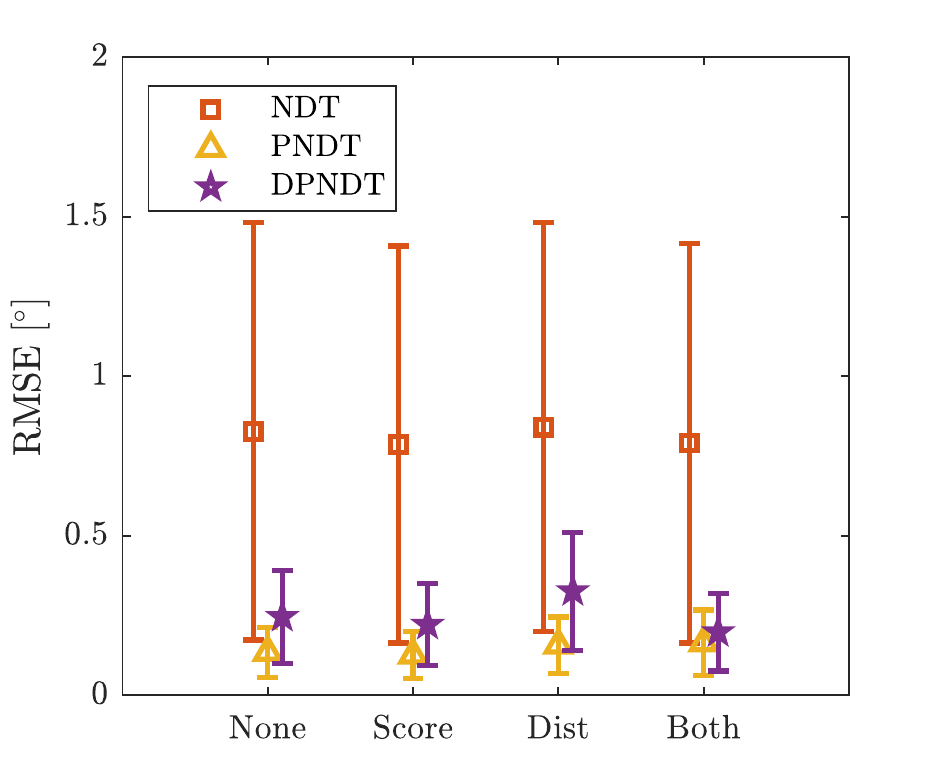}
        \vspace{-2em}
        \caption{}\label{fig:posesHead5}
    \end{subfigure}%         
    \caption{Errors of the pose estimates using the different techniques, (a) the Euclidean distance to the actual positions for cell size 2 meter, (b) the RMSE of the heading estimate for cell size 2 meter, (c) the Euclidean distance to the actual positions for cell size 3 meter, (d) the RMSE of the heading estimate for cell size 3 meter, (e) the Euclidean distance to the actual positions for cell size 5 meter, (f) the RMSE of the heading estimate for cell size 5 meter}\label{fig:resSim}
\end{figure*}

From this starting position, 100 relative poses are randomly generated. These realizations are in line with an almost constant-velocity linear trajectory of a vehicle moving at approximately 15 km/h and a frame rate of 10 Hz. At each pose, a radar frame is generated according to the following steps:
\begin{enumerate}
    \item Extract part of the image around the current pose;
    % \item Add RCS fluctuation to the gray-scale values according to a Swerling III model \cite{swerling}\footnote{Similar to \cite{swerlingcalibration}}
    \item Interpolate the Cartesian image to polar coordinates over a fine grid, restricting the field of view to $\pm 60^\circ$\footnote{This grid is constructed such that the cell size in the range dimension coincides with typical radar range resolution and the cell size in the angle dimension is chosen such as to accommodate a typical Doppler resolution};
    \item Add the Doppler signature in the slow time dimension for each range-angle cell by calculating the Doppler velocity \eqref{eq:Dopp};
    \item Apply the inverse Fast Fourier Transform in the angular dimension to obtain a range-Doppler-spacial frequency element cube. Apply a low-pass filter for the spacial frequencies and down-sample it to create the response of a 12 (3 Tx $\times$ 4 Rx) element array with $\lambda/2$ spacing;
    \item Add thermal noise;
    \item Perform target detection in the range-Doppler domain and apply classical beam-forming for the detected targets to find the corresponding angle. 
    %in the range-Doppler domain;
    % \item Perform classical beamforming for the detected targets to find the corresponding angle
\end{enumerate}
The result is a collection of scans each containing detected targets with corresponding range, angle, Doppler velocity and SNR values. \\

The radar signals are simulated as if they were acquired by a 3 Tx $\times$ 4 Rx automotive radar, capable of measuring at a range resolution of $\Delta R \approx 17$ cm and a Doppler resolution $\Delta v \approx 0.02$ m/s with a maximum measurable range of 40 meters. The measurements are simulated using a signal to noise ratio of 30 dB and a detection threshold of 15 dB.

For the validation of the sensor bias estimation, an artificial bias of $-0.5^\circ$ is added to all measured angles. As mentioned in Section \ref{sec:bias} in order to estimate such a small angular bias, a high Doppler resolution is required, the chosen Doppler resolution of $\Delta v \approx 0.02$ m/s will suffice. 

\begin{table*}[t]
    \centering
    \begin{tabular}{r|cccc|cccc|cccc}
        & \multicolumn{4}{c}{$R = 2$ m} & \multicolumn{4}{c}{$R = 3$ m} & \multicolumn{4}{c}{$R = 5$ m} \\ \cline{2-13}
        & None & Score & Dist & Both & None & Score & Dist & Both & None & Score & Dist & Both \\\hline
    PNDT & 1.10 & 1.10 & 1.11 & 1.11 &          1.06 & 1.06 & 1.06 & 1.05 &     1.10 & 1.10 & 1.11 & 1.11 \\
    DPNDT & 1.49 & 1.42 & 1.51 & 1.49 &         1.47 & 1.47 & 1.45 & 1.47 &     1.69 & 1.50 & 1.64 & 1.44 \\
    \end{tabular}
    \caption{Average execution time of the scan matching algorithms normalized to the average execution time of the conventional NDT}
    \label{tab:timesSim}
\end{table*}

\subsection{Results of the Algorithms Applied to Simulation Data}

The proposed scan-matching techniques are now evaluated on these simulated scans. The resulting pose estimates are compared to the ground truth by calculating the Euclidean distance to the actual position and the root mean square error (RMSE) of the heading, along with their standard deviations, over the 100 frames. Additionally, the execution time is compared and the performance of the sensor bias estimation is examined. 

In order to ensure convergence to the correct minimum, initialization of the estimation vector is needed. This is done by estimating the vehicle velocity by rewriting \eqref{eq:Dopp} into the following form:
\begin{equation}\label{eq:velest}
    \hat{v}_{\mathrm{car},m} = \frac{v_m}{\cos(\theta_m)},
\end{equation}
which results in an estimate of the vehicle velocity based on each point $m$. By taking the median value of these individual values, the vehicle velocity $\hat{v}_\mathrm{car}$ can be estimated. Here $\hat{v}_\mathrm{car}$ is used to initialize $t_x$, by taking into account the frame rate%, and to initialize $v_\mathrm{car} directly$
. 

\subsubsection{Pose Estimates}

Fig. \ref{fig:resSim} shows the average errors of pose estimation over the 100 realisations. In the plots, the performance of the conventional NDT is compared to that of the proposed PNDT and DPNT by showing their respective errors for different grid cell sizes (2 m, 3 m and 5 m) and the different methods of SNR incorporation. 
From these figures it is clear that both the PNDT and the DPNDT show significant improvements over the conventional NDT. The incorporation of the SNR has minor impact on the performance of all techniques, which can be partially explained by the simplicity of the simulated scene. The cell size also seems to have little influence on the results.

\subsubsection{Execution Time}
The average execution times per frame of the PNDT and DPNDT are compared to the execution time of the NDT in Tab. \ref{tab:timesSim}. It shows the average execution time per frame, normalized through division by the average execution times of the NDT for the different simulation set-ups under investigation\footnote{All algorithms were implemented in \textsc{Matlab} with no effort on their efficient/real-time realization. As such, the average execution time of the NDT to match two scans was approximately one second. An efficient implementation similar to the one described in \cite{ndt} can significantly decrease calculation time (in \cite{ndt} the authors claim a frame rate of 97 FPS with the conventional NDT implementation for lidar)}.
%The implementations were not optimized for real-time applications, the average scan matching time per set of frames using the NDT, implemented using , was approximately one second}.
It is seen that the considerable improvement of the PNDT over the NDT comes at only a very small increase in the execution time. 
The DPNDT consumes a factor of 1.4 - 1.7 more computational time than the NDT, slightly more than the 
PNDT, which is expected due to its 3D representation. While not outperforming the PNDT, the DPNDT does show consistency and low errors, which is promising for the additional feature of possible bias correction.

\subsubsection{Sensor Bias Estimation}

\begin{figure}[t]
    \centering
    \begin{subfigure}{\linewidth}
        \figuretitle{Raw estimates}
        \includegraphics[width=\linewidth]{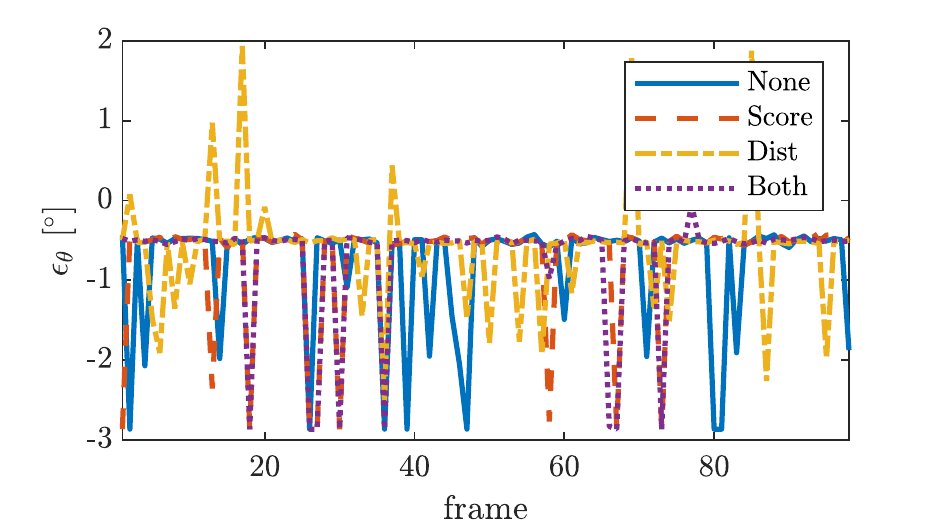}
        \vspace{-1.5em}
        \caption{}\label{fig:biasSimNoMed}
    \end{subfigure}% 
    \\
    \vspace{1em}
    \begin{subfigure}{\linewidth}
        \figuretitle{With median filtering}
        \includegraphics[width=\linewidth]{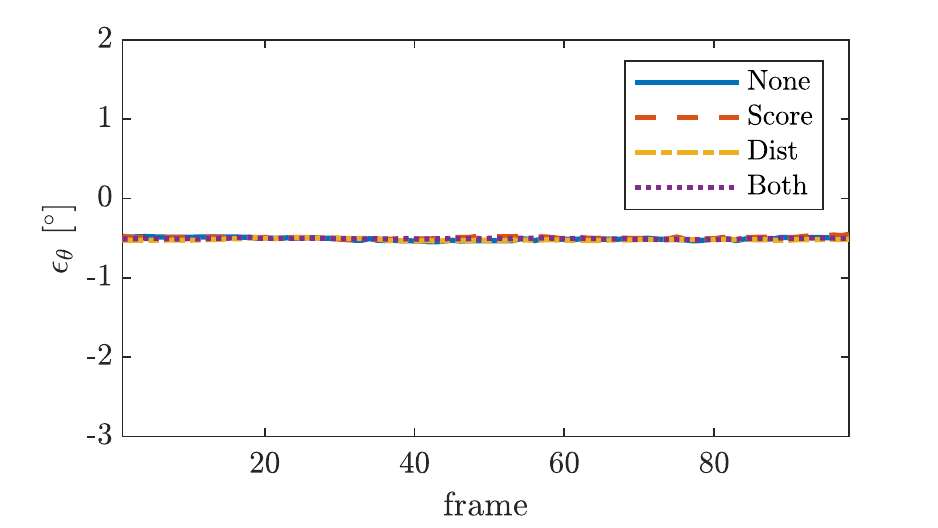}
        \vspace{-1.5em}
        \caption{}\label{fig:biasSimMed}
    \end{subfigure}%    
    \caption{The estimated sensor bias for each of the matched scans, (a) raw estimates, (b) after application of a median filter of order 20}\label{fig:biasSim}
\end{figure}

Fig. \ref{fig:biasSim}, \textit{a} shows the estimated sensor bias for each of the 100 matched frames and for the different SNR incorporation techniques. It shows high consistency around the correct value with some outliers which occur when the algorithm converges to a local minimum. That issue can be resolved by applying e.g. a median filter to the provided estimates. The results after applying a median filter of order 20 are presented in Fig. \ref{fig:biasSim}, \textit{b} and show a very accurate estimation of sensor bias (the variability is less than $0.1^\circ$ after filtering, while the beam-width in the bore-sight is about $10^\circ$). 
The median and average estimated bias over all 100 frames along with the standard deviation (before median filtering) are presented in Tab. \ref{tab:biasSim}. It demonstrates that by taking the median value over all 100 frames, the bias can be estimated with an error of approximately 0.05$^\circ$. The SNR incorporation has minor impact on this result.

% Due to the fact that most outliers are towards one side, the median value is also presented. Based on this, it is seen that when taking the median value and considering the SNR in both the construction of the distribution and the calculation of the score, the bias is estimated with an error of only 1\%.

\begin{table}[t]
    \centering
    \caption{Average value (mean), standard deviation (std) and median of the estimated sensor bias before median filtering for the different SNR incorporation techniques}\label{tab:biasSim}
    \begin{tabular}{r|c|c|c}
                                            & mean [$^\circ$]  & std [$^\circ$]     & median [$^\circ$] \\ \hline
    No SNR incorporation                    & -0.66                   & 0.32                            & -0.55\\
    SNR incorporation in the score          & -0.53                   & 0.50                            & -0.57\\
    SNR incorporation in the distribution   & -0.60                   & 0.88                            & -0.55\\
    SNR incorporation in both               & -0.38                   & 1.30                           & -0.54\\
    \end{tabular}
\end{table}

\section{Experiment}\label{sec:exper}
\begin{figure}[t]
    \centering
    \includegraphics[width=7cm]{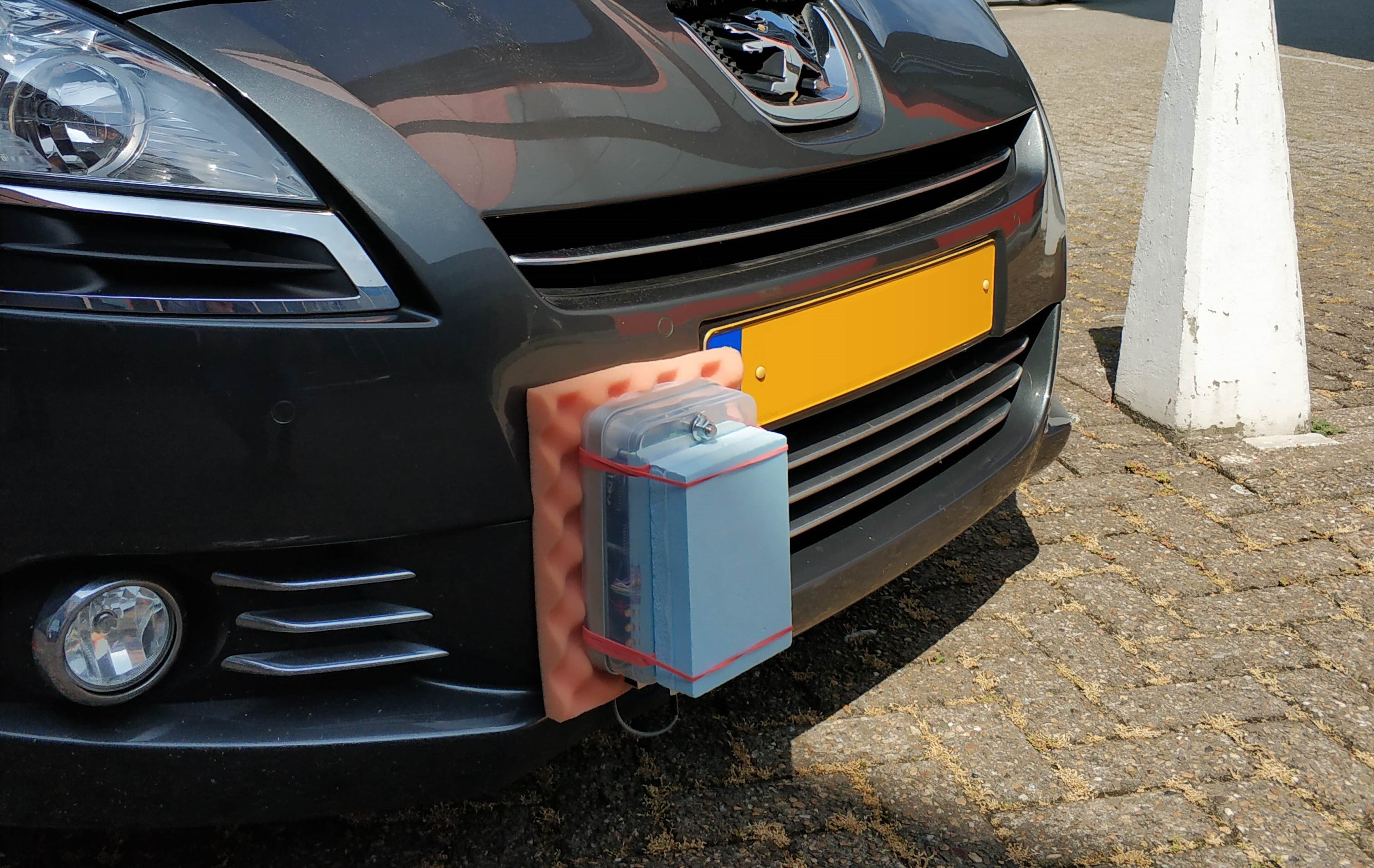}
    \caption{The front of the car with the MIMO radar system attached}
    \label{fig:frontcar}
\end{figure}

% Set-up and results of experiment
For experimental verification, measurements were made using an evaluation board of a commercially available automotive radar chip attached to the front of a car according to the set-up shown in Fig. \ref{fig:frontcar}. The radar system was attached at a slight offset to the center of the car to ensure visibility of the license plate. During the experiment, raw radar data with synchronized GPS locations were recorded, all processing was performed offline. This section describes the experiment set-up along with the results.

\subsection{Set-up of the Experiment}
For collection of the data a commercially available automotive radar chip operating at a frequency of 79 GHz was used. The radar system contains a 3 Tx $\times$ 4 Rx MIMO array operating at a bandwidth of 860 MHz. The angular resolution is approximately $\Delta \theta = 8^\circ$, the range resolution is $\Delta R = 0.17$ m and the Doppler resolution in one frame is $\Delta v = 0.04$ m/s. The radar provides 10 FPS. The maximum measurable range is $R_{max} = 94$ m and the field of view of the radar is $\pm 60^\circ$. During the experiment, the car was driving at  $v_\mathrm{car} \approx 15$ km/h on a quiet road on the TU Delft campus.
The collected data was pre-processed into a range-angle-Doppler data cube after which median detection was performed. In order to filter out any non-stationary targets, the vehicle velocity was estimated for each target within a scan according to \eqref{eq:velest} after which the median value was taken over the points. Using this estimate of the velocity, non-stationary targets were filtered out by discarding any target that did not meet the criterion $|v_m - \hat{v}_{car}\cos(\theta_m)| < 0.5$ m/s.

For performance assessment, the scan matching was applied to 15 seconds of data, which results in 150 data frames.

\begin{figure*}[t]
    \centering
    \begin{subfigure}{0.23\linewidth}
        \figuretitle{No SNR incorporation}
        \includegraphics[width=\linewidth]{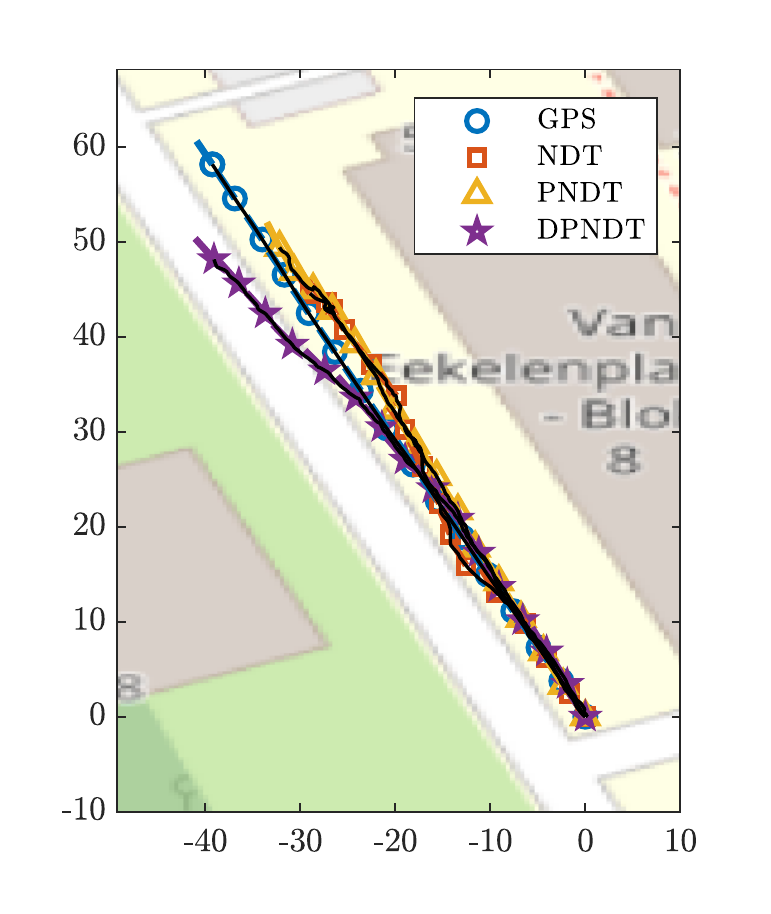}
        \vspace{-2em}
        \caption{}\label{fig:expNone}
    \end{subfigure}% 
    \hspace{1em}
    \begin{subfigure}{0.23\linewidth}
        \figuretitle{SNR in Score}
        \includegraphics[width=\linewidth]{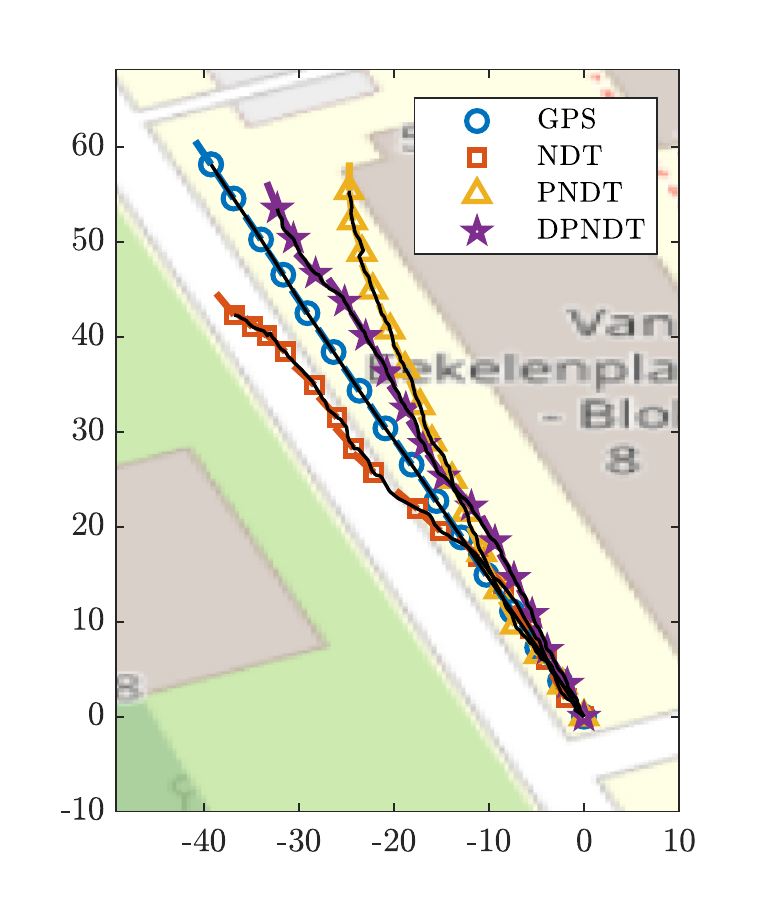}
        \vspace{-2em}
        \caption{}\label{fig:expScore}
    \end{subfigure}%     
    \hspace{1em}
    \begin{subfigure}{0.23\linewidth}
        \figuretitle{SNR in Distribution}
        \includegraphics[width=\linewidth]{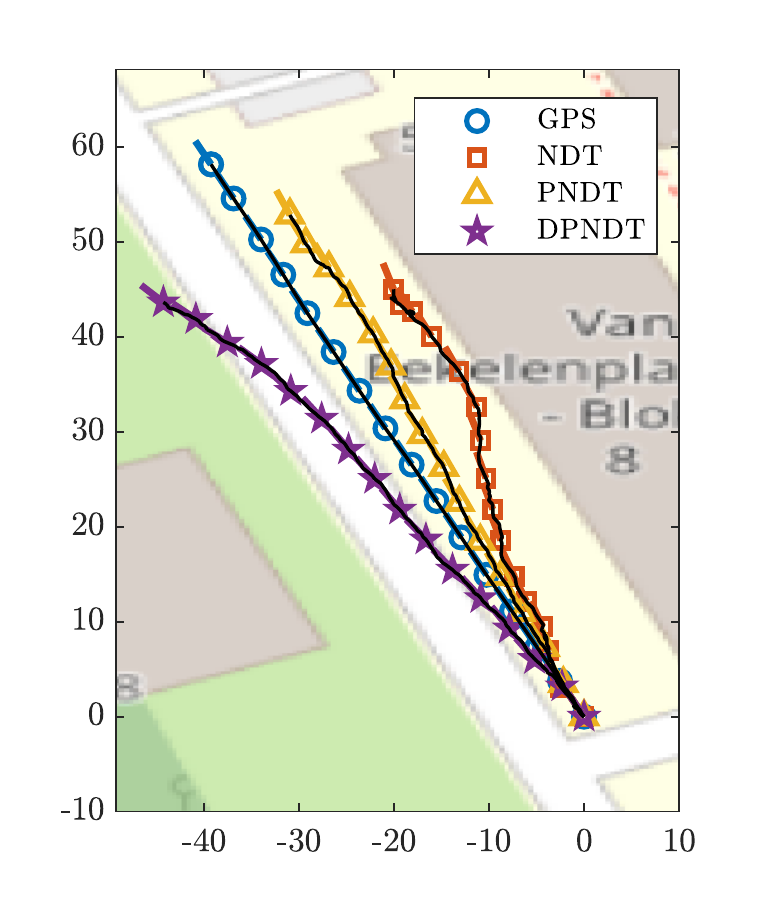}
        \vspace{-2em}
        \caption{}\label{fig:expDist}
    \end{subfigure}%    
    \hspace{1em}
    \begin{subfigure}{0.23\linewidth}
        \figuretitle{SNR in Both}
        \includegraphics[width=\linewidth]{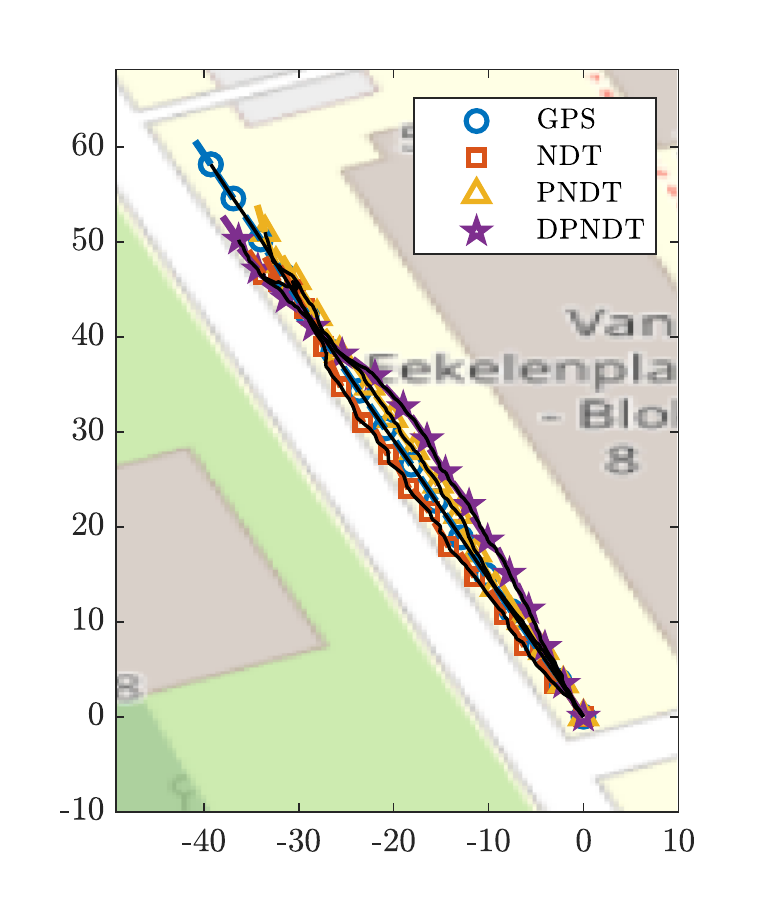}
        \vspace{-2em}
        \caption{}\label{fig:expBoth}
    \end{subfigure}%     
    \caption{Reconstructed trajectories for each of the investigated techniques, using the GPS as a reference, (a) not taking into account the SNR measurements, (b) taking into account the SNR measurements in the construction of the distribution, (c) taking into account the SNR measurements in the calculation of the score, (d) taking into account the SNR measurements in both the distribution and the score}
    \label{fig:resExp}
\end{figure*}

\subsection{Results of the Algorithms Applied to Experimental Data}
Similarly to the above, the estimation vector is initialized using an estimate of the vehicle velocity according to \eqref{eq:velest} and a cell size of 0.5 meters is chosen. As mentioned before, the GPS data does not present a reliable ground truth. To assess the performance of the different techniques, the individual pose estimates are combined into a trajectory after crude filtering of outliers, which is then drawn on top of a map of the surroundings, using the GPS data as a reference. The filtering is performed by discarding estimates that are not reasonable for the assumed vehicle model, removing estimates of which the translation in the $x$-direction is more than 0.75 m, the translation in the $y$-direction is more than 0.2 m, and the change in heading is more than $10^\circ$. These are very reasonable assumptions for the vehicle's velocity at a framerate of 10 Hz, even for a nonlinear vehicle motion model. Discarded estimates are replaced by a vector containing estimates $t_x$ and $t_y$ of the previous frame and a heading $\phi$ of zero. The results of this approach are given in Fig. \ref{fig:resExp}. In this representation, the traversed trajectories are indicated by the black lines, the shapes and bars denote the estimated position and heading, respectively, at time intervals of 10 frames (1 second). The offset from the road in the reconstructed trajectory is due to the starting point being chosen according to GPS data. Immediately it stands out that the techniques underestimate the traversed distance in the $t_x$-direction as compared to the GPS. A possible reason for that is a synchronization error between the GPS and radar measurements: a GPS receiver providing 1 FPS was used and the data acquisition was done in a home-made software not optimized for real-time application, which resulted in the fact that the number of frames with the same GPS stamp had a jitter between 8 and 12. 
When looking at the distance traversed according to the GPS within any given time frame, the velocity needed to accomplish this is higher than the velocity that was driven during the experiment. This observation is corroborated by the fact that the maximum measured Doppler velocities are consistently approximately 0.5 m/s lower than needed to travel the distance provided by the GPS data.

\begin{figure}[t]
    \centering
    \begin{subfigure}{\linewidth}
        \figuretitle{Without artificial bias}
        \includegraphics[width=\linewidth]{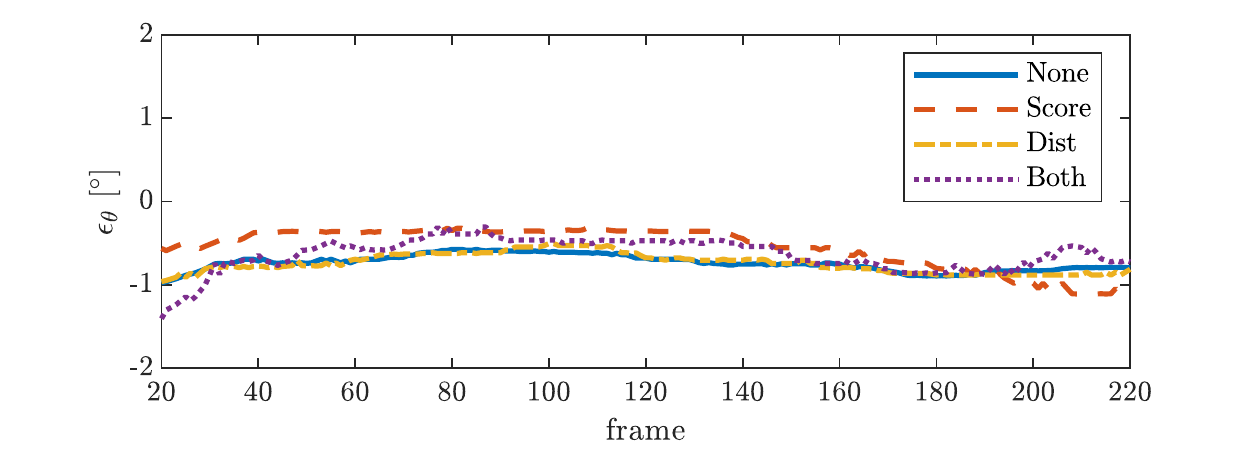}
        \vspace{-1.5em}
        \caption{}\label{fig:biasExpBeforeMed}
    \end{subfigure}% 
    \\
    \vspace{1em}
    \begin{subfigure}{\linewidth}
        \figuretitle{With artificial bias}
        \includegraphics[width=\linewidth]{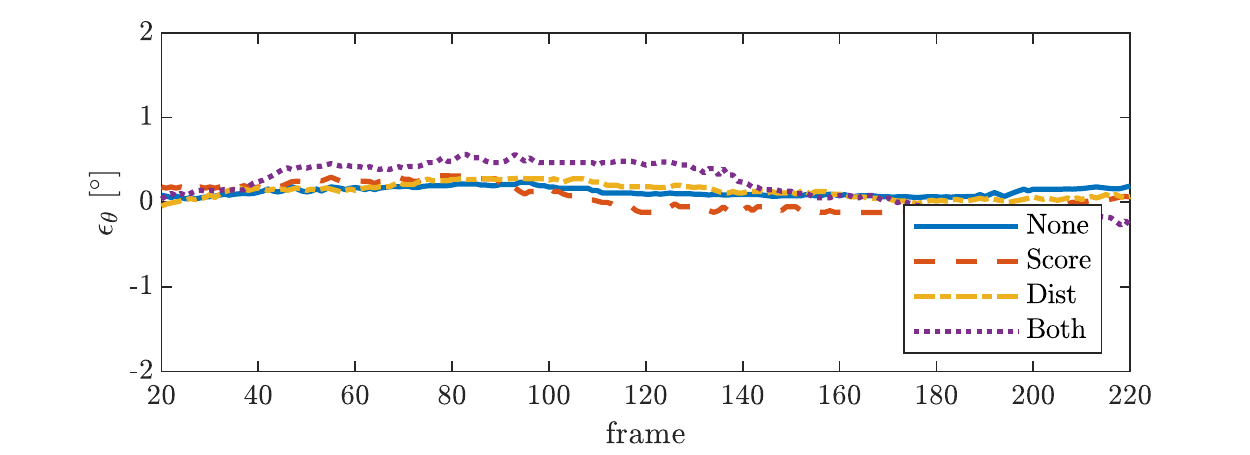}
        \vspace{-1.5em}
        \caption{}\label{fig:biasExpAfterMed}
    \end{subfigure}%    
    \\
    \vspace{1em}
    \begin{subfigure}{\linewidth}
        \figuretitle{Difference}
        \includegraphics[width=\linewidth]{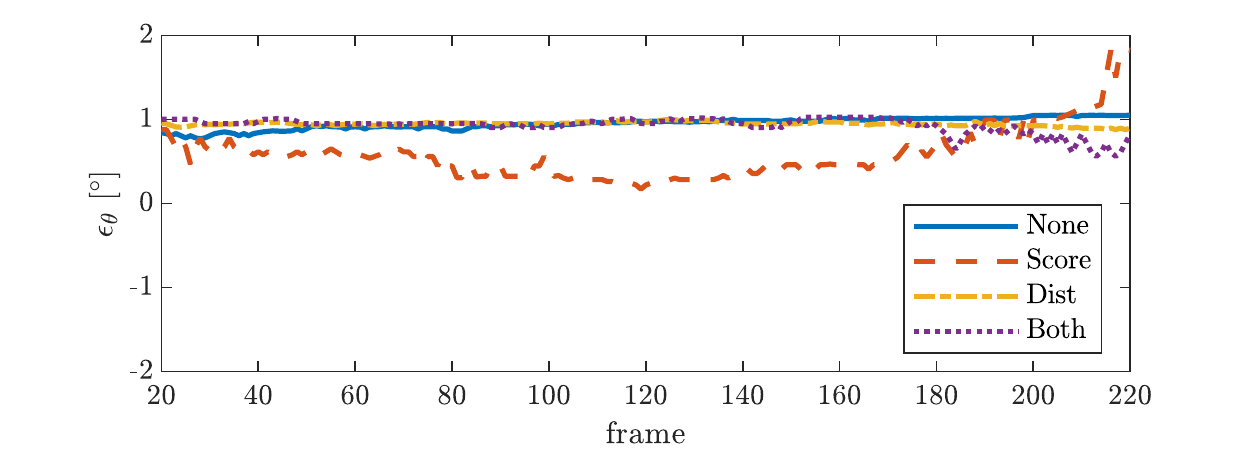}
        \vspace{-1.5em}
        \caption{}\label{fig:biasDiffExpMed}
    \end{subfigure}%    
    \caption{The estimated sensor bias for each of the matched scans after application of a median filter of order 100, (a) before addition of the artificial bias, (b) after addition of the artificial bias, (c) the difference between the estimates with and without artificial bias}\label{fig:biasExpMed}
\end{figure}

\subsubsection{Pose Estimation}
The main problem that can be seen with the unaltered NDT is the poor performance regarding the estimation of the traversed distance in the $t_x$-direction. It is also seen that the estimation errors within the trajectory propagate through to a large offset at the end of the trajectory, which is especially clear from the trajectory of the DPNDT in Fig. \ref{fig:resExp}, \textit{c}. This makes this way of presenting the results somewhat suboptimal. Overall, it is clear that the estimates using the DPNDT construct the smoothest trajectories, with very little sporadic changes. In each case, the DPNDT underestimates the traversed distance in the direction of $t_x$ by $7.5-8.5$ m, which is in line with the higher velocity according to the GPS data of approximately 0.5 m/s over 15 seconds (150 consecutive frames at 10 Hz). In this case, the incorporation of SNR does seem to have an effect; especially in the case where the SNR is taken into account in both the score and the distribution, the trajectories for each technique follow a fairly straight path.

% It is clear that the unaltered NDT shows poor performance overall, especially underestimating the traversed distance in the $t_x$-direction significantly compared to the proposed techniques. The incorporation of SNR measurements does seem to have some influence here. It is clearly seen that the trajectories constructed for each technique suffer quite significant curvature if no SNR incorporation is included. The PNDT seems to benefit most from the incorporation in the score, whereas the DPNDT shows best performance with incorporation of SNR measurements in both the distribution and the score. For each of these best performing scenarios, the traversed distance is underestimated compared to the GPS by approximately 8 and 9 meters, respectively, which is in line with the higher velocity according to the GPS data of approximately 0.5-0.8 m/s over 15 seconds (150 consecutive frames at 10 Hz).

\subsubsection{Execution Time}
The execution times of the PNDT and DPNDT are presented in Tab. \ref{tab:timesExp}. Again, this table shows the average execution time normalized to that of the NDT. It is clear that in this case the DPNDT suffers from significant increase of the computational complexity, resulting in much longer execution times.

\begin{table}[t]
    \centering
    \begin{tabular}{r|cccc}
        & None & Score & Dist & Both \\\hline
    PNDT & 1.08 & 1.14 & 0.94 & 1.08 \\
    DPNDT & 6.20 & 7.70 & 5.59 & 8.10 \\
    \end{tabular}
    \caption{Average execution time of the scan matching algorithms normalized to the average execution time of the conventional NDT}
    \label{tab:timesExp}
\end{table}

\begin{table*}[t]
    \centering
    \caption{Average value, standard deviation and median of the estimated sensor bias for the different SNR incorporation techniques}\label{tab:biasExp}
    \begin{tabular}{r|c|c|c}
        & w/o art. bias [$^\circ$] & w/ art. bias [$^\circ$] & Difference [$^\circ$] \\\hline
    No SNR incorporation & -0.75 & 0.11 & 0.97\\
    SNR incorporation in the score & -0.51 & 0.01 & 0.55 \\
    SNR incorporation in the distribution & -0.79 & 0.12 & 0.95 \\
    SNR incorporation in both & -0.69 & 0.15 & 0.95 \\
    \end{tabular}
\end{table*}

\subsubsection{Sensor Bias Estimation}
During the gathering of experimental data, the radar system was attached to the front of the car at a slight offset from the center to ensure visibility of the license plate. Because of the curvature of the bumper this resulted in a bias in the angular measurements. This bias, however, was not measured. Moreover, the last full antenna array calibration in an anechoic chamber was performed approximately one year before this measurement campaign. Therefore, the ground truth of the angular bias was not accessible.
To validate the performance of the sensor bias estimation technique, an artificial bias of $1^\circ$ was added to each of the experimental data frames, similar to the approach in the verification on simulations. The sensor bias is then estimated for the case with and without the addition of this artificial bias and the difference is compared to the chosen artificial bias.

For the estimation of the sensor bias, 200 consecutive frames were used. This larger amount of frames was chosen since the sensor bias estimate showed a large amount of outliers. The considered frames are frame 20 through 220, the first 20 frames are discarded since the vehicle was stationary at that time. Fig. \ref{fig:biasExpMed} shows the estimated sensor bias per frame before and after addition of the artificial bias, after application of a median filter of order 100 (10 seconds of data). The 200 consecutive, unfiltered estimates before the addition of the artificial bias are subtracted from the estimates after the addition of the artificial bias. To this difference a similar median filter is applied. The results are shown in Fig. \ref{fig:biasExpMed}, \textit{c}.

Tab. \ref{tab:biasExp} shows the median values of the estimated sensor bias before and after addition of the artificial bias, along with the median of their differences. The median estimation of the bias in the default setting shows that the radar measurements suffered a bias of $-0.5$ to $-0.8^\circ$ from the heading direction of the car, which can be a reasonable consequence of the installation on the bumper and of a possible drift of the calibration coefficients. The results with adding artificial bias demonstrate that the added difference can be estimated with approximately $0.03-0.05^\circ$ accuracy. In the case where only the score is weighted according to the SNR, the bias estimate suffers a considerable offset, which is also seen when looking at Fig. \ref{fig:biasExpMed}, \textit{c}.

\section{Conclusion}\label{sec:conc}

In this paper, we have addressed the problem of accurate vehicle localization using high-resolution radar measurements. 
The improvements over the Normal Distributions Transform (NDT) --- an established technique for vehicle localization --- are brought by solving it in the natural for the radar sensor polar coordinate system, introduced by the Polar Normal Distributions Transform (PNDT). 
The technique is further extended to the Doppler PNDT (DPNDT), which allows for the first time to incorporate Doppler velocity measurements in a scan-matching algorithm. 
Furthermore, the impact of target SNR on the localization performance is accounted for with the additional modification of the cost function and distribution, which makes the algorithms more resistant to target RCS fluctuations, compared to the existing scan-matching techniques. Simulation results and experimental data processing demonstrate a significant improvement of the proposed techniques (PNDT and DPNDT) --- by a factor of 3 to 5 in the MSE of localization --- over the conventional NDT. This comes with only a small increase (about 10-70\%) of the computational load. Real data processing shows that the incorporation of SNR in both the score and the distribution is a beneficial approach for the NDT, PNDT and DPNDT.

Moreover, we demonstrated that the explicit dependency of the Doppler measurements and the target angle of incidence can be used to correct the angular bias of the radar. Simulation results and real data processing indicated that this bias can be estimated with $~0.1^o$ accuracy by processing just a few seconds of data in real-time operation. 

\section*{Acknowledgment}

The authors thank Fred van der Zwan, Peter Swart and Pascal Aubry for their help with the radar measurements and Hans Driessen for his valuable  feedback on this study.

\section*{Appendix A. Elements of gradient and Hessian matrices of PNDT}
\label{Appendix_patial_derivatives}
% \hfill

\subsection{The First Order Partial Derivatives}
\textsc{Range:}
\begin{gather*}
    r'_m = \smashedsqrt{\underbrace{[t_x + r_m \cos (\phi + \theta_m)]^2 + [t_y + r_m \sin (\phi + \theta_m)]^2}_{\text{\normalfont $\alpha$}}}\\
    \frac{\partial r'_m}{\partial t_x} = \frac{[t_x+r_m\cos(\phi + \theta_m)]}{\sqrt{\alpha}} \\
    \frac{\partial r'_m}{\partial t_y} = \frac{[t_y+r_m\sin(\phi + \theta_m)]}{\sqrt{\alpha}} \\
    \frac{\partial r'_m}{\partial \phi} = \frac{r_m[t_y\cos(\phi + \theta_m)-t_x\sin(\phi + \theta_m)]}{\sqrt{\alpha}}
\end{gather*}

\hfill

\textsc{Angle:}
\begin{gather*}
        \theta'_m = \atantwo[t_y + r_m \sin (\phi + \theta_m), t_x + r_m \cos (\phi + \theta_m)] \\
        = \arctan\Bigg[\underbrace{\frac{t_y + r_m \sin (\phi + \theta_m)}{t_x + r_m \cos (\phi + \theta_m)}}_{\text{\normalfont $\beta$}}\Bigg] + C,
\end{gather*}
with $C$ a constant depending on the quadrant.
Using notation $\mathbf{p} = [t_x, t_y, \phi]$, we can write:
\begin{gather*}
    \frac{\partial \theta'_m}{\partial p_i} = \frac{1}{1+\beta^2} \frac{\partial \beta}{\partial p_i}\\
    \frac{\partial \beta}{\partial t_x} =  \frac{-t_y - r_m\sin(\phi + \theta_m)}{\left[t_x + r_m\cos(\phi + \theta_m)\right]^2}\\
    \frac{\partial \beta}{\partial t_y} =  \frac{1}{t_x + r_m\cos(\phi + \theta_m)} \\
    \frac{\partial \beta}{\partial \phi} =  \frac{r_m^2 + t_x r_m \cos(\phi + \theta_m) + t_y r_m \sin(\phi + \theta_m)}{\left[t_x + r_m\cos(\phi + \theta_m)\right]^2}
\end{gather*}
where $i \in \{1,2,3\}$.
%\begin{gather*}
%    \frac{\partial \theta'_m}{\partial t_x} = \frac{1}{1+\beta^2} \frac{\partial \beta}{\partial t_x} = \frac{1}{1+\beta^2} \frac{-t_y - r_m\sin(\phi + \theta_m)}{\left[t_x + r_m\cos(\phi + \theta_m)\right]^2}\\
%   \frac{\partial \theta'_m}{\partial t_y} = \frac{1}{1+\beta^2} \frac{1}{t_x + r_m\cos(\phi + \theta_m)} \\
%    \frac{\partial \theta'_m}{\partial \phi} = \frac{1}{1+\beta^2} \frac{r_m^2 + t_x r_m \cos(\phi + \theta_m) + t_y r_m \sin(\phi + \theta_m)}{\left[t_x + r_m\cos(\phi + \theta_m)\right]^2}
%\end{gather*}

\subsection{The Second Order Partial derivatives}\mbox{}\\
\textsc{Range:}

%First Derivative to $t_x$:
\begin{gather*}
 %   \frac{\partial r'_m}{\partial t_x} = \frac{[t_x+r_m\cos(\phi + \theta_m)]}{\sqrt{\alpha}} \\
    \frac{\partial^2 r'_m}{\partial t_x^2} = \frac{\sqrt{a}-[t_x+r_m\cos(\phi + \theta_m)]\frac{\partial r'_m}{\partial t_x}}{\alpha}\\
    \frac{\partial^2 r'_m}{\partial t_x \partial t_y} = -\frac{[t_x+r_m\cos(\phi + \theta_m)]}{\alpha} \frac{\partial r'_m}{\partial t_y}\\
    \frac{\partial^2 r'_m}{\partial t_x \partial \phi} = -\frac{r_m\sqrt{a}\sin(\phi + \theta_m)+[t_x+r_m\cos(\phi + \theta_m)]\frac{\partial r'_m}{\partial \phi}}{\alpha} \\
%\end{gather*}
%\hfill
%First Derivative to $t_y$:
%\begin{gather*}
%    \frac{\partial r'_m}{\partial t_y} = \frac{[t_y+r_m\sin(\phi + \theta_m)]}{\sqrt{\alpha}} \\
    \frac{\partial^2 r'_m}{\partial t_y^2} = \frac{\sqrt{a}-[t_y+r_m\sin(\phi + \theta_m)]\frac{\partial r'_m}{\partial t_y}}{\alpha}\\
    \frac{\partial^2 r'_m}{\partial t_y \partial \phi} = \frac{r_m\sqrt{a}\cos(\phi + \theta_m)-[t_y+r_m\sin(\phi + \theta_m)]\frac{\partial r'_m}{\partial \phi}}{\alpha} \\
%\end{gather*}
%\hfill
%First Derivative to $\phi$:
%\begin{gather*}
%    \frac{\partial r'_m}{\partial \phi} = \frac{r_m[t_y\cos(\phi + \theta_m)-t_x\sin(\phi + \theta_m)]}{\sqrt{\alpha}} \\
    \begin{multlined}
        \frac{\partial^2 r'_m}{\partial \phi^2} = \frac{-r_m \sqrt{a}[t_x \cos(\phi + \theta_m) + t_y \sin(\phi + \theta_m)]}{\alpha} \\
        + \frac{r_m [t_x\sin(\phi + \theta_m) - t_y \cos(\phi + \theta_m)]\frac{\partial r'_m}{\partial \phi}}{\alpha}
    \end{multlined}
\end{gather*}

\hfill

\textsc{Angle:}

Using the auxiliary variable $\beta$ defined above, we can write:
%Recall the statement made previously:
%\begin{gather*}
%\frac{\partial \theta'_m}{\partial t_x} = \frac{1}{1+\beta^2} \frac{\partial \beta}{\partial t_x}
%\end{gather*}
\begin{align*}
%     \frac{\partial^2 \theta'_m}{\partial p_i^2} 
%    &= \left[\frac{1}{1+\beta^2} \frac{\partial^2 \beta}{\partial p_i^2} -\frac{2 \beta}{(1 + \beta^2)^2} \left(\frac{\partial \beta}{\partial p_i}\right)^2 \right], \\
    \frac{\partial^2 \theta'_m}{\partial p_i \partial p_j} 
    &= \left[\frac{1}{1+\beta^2} \frac{\partial^2 \beta}{\partial p_i \partial p_j} -\frac{2 \beta}{(1 + \beta^2)^2} \frac{\partial \beta}{\partial p_i} \frac{\partial \beta}{\partial p_j} \right],
\end{align*}
where  $i, j \in \{1,2,3\}$.
%This results in the following observations:
%\begin{align*}
%    \frac{\partial^2 \theta'_m}{\partial t_x^2} &= \frac{\partial}{\partial t_x}\left( \frac{1}{1+\beta^2}\frac{\partial \beta}{\partial t_x} \right) \\
%    &= \left[\frac{1}{1+\beta^2} \frac{\partial^2 \beta}{\partial t_x^2} -\frac{2 \beta}{(1 + \beta^2)^2} \left(\frac{\partial \beta}{\partial t_x}\right)^2 \right], \\
 %   \frac{\partial^2 \theta'_m}{\partial t_x \partial t_y} &= \frac{\partial}{\partial t_y}\left( \frac{1}{1+\beta^2}\frac{\partial \beta}{\partial t_x} \right) \\
  %  &= \left[\frac{1}{1+\beta^2} \frac{\partial^2 \beta}{\partial t_x \partial t_y} -\frac{2 \beta}{(1 + \beta^2)^2} \frac{\partial \beta}{\partial t_y} \frac{\partial \beta}{\partial t_x} \right] \textrm{, etc.}
%\end{align*}

% Similar expressions can be found for each of the other second order derivatives. The following parts discuss the second order partial derivatives of $\beta$.

% \hfill
The second-order partial derivatives of $\beta$ then given by:
%First Derivative to $t_x$:
\begin{gather*}
%    \frac{\partial \beta}{\partial t_x} = \frac{-t_y - r_m\sin(\phi + \theta_m)}{\left[t_x + r_m\cos(\phi + \theta_m)\right]^2} \\
    \frac{\partial^2 \beta}{\partial t_x^2} = 2 \frac{t_y + r_m \sin(\phi+\theta_m)}{\left[t_x + r_m\cos(\phi + \theta_m)\right]^3}\\
    \frac{\partial^2 \beta}{\partial t_x \partial t_y} = \frac{-1}{\left[t_x + r_m\cos(\phi + \theta_m)\right]^2}\\
    \begin{multlined}
        \frac{\partial^2 \beta}{\partial t_x \partial \phi} = \frac{- r_m^2[\sin^2(\phi + \theta_m)+1] - 2 t_y r_m\sin(\phi + \theta_m)}{\left[t_x + r_m\cos(\phi + \theta_m)\right]^3} \\
        - \frac{t_x r_m\cos(\phi + \theta_m)}{\left[t_x + r_m\cos(\phi + \theta_m)\right]^3}
    \end{multlined} \\
%\end{gather*}
%First Derivative to $t_y$:
%\begin{gather*}
%    \frac{\partial \beta}{\partial t_y} = \frac{1}{t_x + r_m\cos(\phi + \theta_m)} \\
        \frac{\partial^2 \beta}{\partial t_y^2} = 0 \\
        \frac{\partial^2 \beta}{\partial t_y \partial \phi} = \frac{r_m\sin(\phi + \theta_m)}{\left[t_x + r_m\cos(\phi + \theta_m)\right]^2} \\
%\end{gather*}
%\hfill
%First Derivative to $\phi$:
%\begin{gather*}
%    \frac{\partial \beta}{\partial \phi} = \frac{r_m^2 + t_x r_m \cos(\phi + \theta_m) + t_y r_m \sin(\phi + \theta_m)}{\left[t_x + r_m\cos(\phi + \theta_m)\right]^2} \\
    \begin{multlined}
        \frac{\partial^2 \beta}{\partial \phi^2} = \frac{2 r_m^3 \sin(\phi + \theta_m) + t_y r_m^2 [1+ \sin^2(\phi + \theta_m)]}{[t_x + r_m\cos(\phi + \theta_m)]^3} \\
        + \frac{t_x r_m [t_y \cos(\phi + \theta_m) - t_x \sin(\phi + \theta_m)]}{[t_x + r_m\cos(\phi + \theta_m)]^3} \\
        + \frac{t_x r_m^2 \sin(\phi + \theta_m) \cos(\phi + \theta_m)}{[t_x + r_m\cos(\phi + \theta_m)]^3}
    \end{multlined}
\end{gather*}

\section*{Appendix B. Elements of gradient and Hessian matrices of DPNDT}
% %
Recall:
\begin{equation*}
    \begin{aligned}
        v'_m &= f\underbrace{\sqrt{t_x^2+t_y^2}}_{\gamma}\underbrace{\cos(\theta'_m)}_{\delta}
    \end{aligned}
\end{equation*}
with $\theta'_m$ being defined in \eqref{eq:mappingeqsPNDT}.

\subsection{The First Order Partial Derivatives}
\begin{gather*}
    % v'_m = v_\mathrm{car}\cos(\theta'_m)\\
    \frac{\partial \gamma}{\partial t_x} = \frac{t_x}{\gamma} \\
    \frac{\partial \gamma}{\partial t_y} = \frac{t_y}{\gamma} \\
    \frac{\partial \gamma}{\partial \phi} = 0
\end{gather*}
\begin{gather*}
    % v'_m = v_\mathrm{car}\cos(\theta'_m)\\
    \frac{\partial \delta}{\partial t_x} = -\sin(\theta'_m) \frac{\partial \theta'_m}{\partial t_x} \\ 
    \frac{\partial \delta}{\partial t_y} = -\sin(\theta'_m) \frac{\partial \theta'_m}{\partial t_y} \\
    \frac{\partial \delta}{\partial \phi} = -\sin(\theta'_m) \frac{\partial \theta'_m}{\partial \phi}
\end{gather*}
\begin{gather*}
    % v'_m = v_\mathrm{car}\cos(\theta'_m)\\
    \frac{\partial v'_m}{\partial t_x} = f \left(\frac{\partial \gamma}{\partial t_x} \delta + \gamma \frac{\partial \delta}{\partial t_x}\right) \\
    \frac{\partial v'_m}{\partial t_y} = f \left(\frac{\partial \gamma}{\partial t_y} \delta + \gamma \frac{\partial \delta}{\partial t_y}\right) \\
    \frac{\partial v'_m}{\partial \phi} = f \left(\gamma \frac{\partial \delta}{\partial \phi}\right) \\
\end{gather*}

\subsection{The Second Order Partial derivatives}
\begin{gather*}
    \frac{\partial^2 \gamma}{\partial t_x^2} = \frac{t_y^2}{\gamma^3} \\
    \frac{\partial^2 \gamma}{\partial t_x \partial t_y} = \frac{-t_x t_y}{\gamma^3} \\
    \frac{\partial^2 \gamma}{\partial t_x \partial \phi} = 0\\
\end{gather*}
\begin{gather*}
    \frac{\partial^2 \gamma}{\partial t_y^2} = \frac{t_x^2}{\gamma^3} \\
    \frac{\partial^2 \gamma}{\partial t_x \partial \phi} = 0\\
\end{gather*}
\begin{gather*}
    \frac{\partial^2 \gamma}{\partial \phi^2} = 0 \\
\end{gather*}
%
%----------------------------
\begin{gather*}
    \frac{\partial^2 \delta}{\partial t_x^2} = -\delta\left(\frac{\partial \theta'_m}{\partial t_x}\right)^2 - \sin (\theta'_m)\frac{\partial^2 \theta'_m}{\partial t_x^2} \\
    \frac{\partial^2 \delta}{\partial t_x \partial t_y} = -\delta\frac{\partial \theta'_m}{\partial t_x}\frac{\partial \theta'_m}{\partial t_y} - \sin (\theta'_m)\frac{\partial^2 \theta'_m}{\partial t_x \partial t_y} \\
    \frac{\partial^2 \delta}{\partial t_x \partial \phi} = -\delta\frac{\partial \theta'_m}{\partial t_x}\frac{\partial \theta'_m}{\partial \phi} - \sin (\theta'_m)\frac{\partial^2 \theta'_m}{\partial t_x \partial \phi}\\
\end{gather*}
\begin{gather*}
    \frac{\partial^2 \delta}{\partial t_y^2} = -\delta\left(\frac{\partial \theta'_m}{\partial t_y}\right)^2 - \sin (\theta'_m)\frac{\partial^2 \theta'_m}{\partial t_y^2} \\
    \frac{\partial^2 \delta}{\partial t_y \partial \phi} = -\delta\frac{\partial \theta'_m}{\partial t_y}\frac{\partial \theta'_m}{\partial \phi} - \sin (\theta'_m)\frac{\partial^2 \theta'_m}{\partial t_y \partial \phi} \\
\end{gather*}
\begin{gather*}
    \frac{\partial^2 \delta}{\partial \phi^2} = -\delta\left(\frac{\partial \theta'_m}{\partial \phi}\right)^2 - \sin (\theta'_m)\frac{\partial^2 \theta'_m}{\partial \phi^2} \\
\end{gather*}
%
% First Derivative to $t_x$:
\begin{gather*}
     %\frac{\partial v'_m}{\partial t_x} = -v_\mathrm{car}\sin(\theta'_m) \frac{\partial \theta'_m}{\partial t_x} \\
    \frac{\partial^2 v'_m}{\partial t_x^2} = f \left(\frac{\partial^2 \gamma}{\partial t_x^2} \delta + \frac{\partial^2 \delta}{\partial t_x^2} \gamma + 2 \frac{\partial \gamma}{\partial t_x} \frac{\partial \delta}{\partial t_x} \right) \\
    \begin{multlined}
    \frac{\partial^2 v'_m}{\partial t_x \partial t_y} = f \bigg(\frac{\partial^2 \gamma}{\partial t_x \partial t_y} \delta + \frac{\partial^2 \delta}{\partial t_x \partial t_y} \gamma \\ 
    + \frac{\partial \gamma}{\partial t_x} \frac{\partial \delta}{\partial t_y} + \frac{\partial \gamma}{\partial t_y} \frac{\partial \delta}{\partial t_x} \bigg) 
    \end{multlined} \\
    \frac{\partial^2 v'_m}{\partial t_x \partial \phi} = f \left(\frac{\partial^2 \delta}{\partial t_x \partial \phi} \gamma + \frac{\partial \gamma}{\partial t_x} \frac{\partial \delta}{\partial \phi}\right)
\end{gather*}
% \hfill
%  First Derivative to $t_y$:
\begin{gather*}
%     \frac{\partial v'_m}{\partial t_y} = -v_\mathrm{car}\sin(\theta'_m) \frac{\partial \theta'_m}{\partial t_y} \\
    \frac{\partial^2 v'_m}{\partial t_y^2} = f \left(\frac{\partial^2 \gamma}{\partial t_y^2} \delta + \frac{\partial^2 \delta}{\partial t_y^2} \gamma + 2 \frac{\partial \gamma}{\partial t_y} \frac{\partial \delta}{\partial t_y} \right) \\
    \frac{\partial^2 v'_m}{\partial t_y \partial \phi} = f \left(\frac{\partial^2 \delta}{\partial t_y \partial \phi} \gamma + \frac{\partial \gamma}{\partial t_y} \frac{\partial \delta}{\partial \phi}\right)
\end{gather*}
% \hfill
% First Derivative to $\phi$:
\begin{gather*}
%     \frac{\partial v'_m}{\partial \phi} = -v_\mathrm{car}\sin(\theta'_m) \frac{\partial \theta'_m}{\partial \phi} \\
    \frac{\partial^2 v'_m}{\partial \phi^2} = f \frac{\partial^2 \delta}{\partial \phi^2} \gamma
\end{gather*}

%%%%%%%%%%%%%%%%%%%%%%%%%%%%%%%%%%%%
\section*{Appendix C. Elements of gradient and Hessian matrices of DPNDT with joint sensor bias estimation}
\label{appendix_bias}
%%%%%%%%%%%%%%%%%%%%%%%%%%%%%%%%%%%
\setcounter{subsection}{0}

\subsection{The First Order Partial Derivatives}
\begin{gather*}
    \frac{\partial r'_m}{\partial \epsilon_{\theta}} = - \frac{\partial r'_m}{\partial \phi} \\
    \frac{\partial \theta'_m}{\partial \epsilon_{\theta}} = - \frac{\partial \theta'_m}{\partial \phi} + 1\\
    \frac{\partial v'_m}{\partial \epsilon_{\theta}} = - \frac{\partial v'_m}{\partial \phi}
\end{gather*}
\subsection{The Second Order Partial Derivatives}\mbox{}\\
\textsc{Range:}
\begin{gather*}
    \frac{\partial^2 r'_m}{\partial t_x \partial \epsilon_{\theta}} = - \frac{\partial^2 r'_m}{\partial t_x \partial \phi} \\
    \frac{\partial^2 r'_m}{\partial t_y \partial \epsilon_{\theta}} = - \frac{\partial^2 r'_m}{\partial t_y \partial \phi} \\
    \frac{\partial^2 r'_m}{\partial \phi \partial \epsilon_{\theta}} = - \frac{\partial^2 r'_m}{\partial \phi^2} \\
    \frac{\partial^2 r'_m}{\partial \epsilon_{\theta}^2} = \frac{\partial^2 r'_m}{\partial \phi^2}
\end{gather*}
\textsc{Angle:}
\begin{gather*}
    \frac{\partial^2 \theta'_m}{\partial t_x \partial \epsilon_{\theta}} = - \frac{\partial^2 \theta'_m}{\partial t_x \partial \phi} \\
    \frac{\partial^2 \theta'_m}{\partial t_y \partial \epsilon_{\theta}} = - \frac{\partial^2 \theta'_m}{\partial t_y \partial \phi} \\
    \frac{\partial^2 \theta'_m}{\partial \phi \partial \epsilon_{\theta}} = - \frac{\partial^2 \theta'_m}{\partial \phi^2} \\
    \frac{\partial^2 \theta'_m}{\partial \epsilon_{\theta}^2} = \frac{\partial^2 \theta'_m}{\partial \phi^2}
\end{gather*}
\textsc{Velocity:}
\begin{gather*}
    \frac{\partial^2 v'_m}{\partial t_x \partial \epsilon_{\theta}} = - \frac{\partial^2 v'_m}{\partial t_x \partial \phi} \\
    \frac{\partial^2 v'_m}{\partial t_y \partial \epsilon_{\theta}} = - \frac{\partial^2 v'_m}{\partial t_y \partial \phi} \\
    \frac{\partial^2 v'_m}{\partial \phi \partial \epsilon_{\theta}} = - \frac{\partial^2 v'_m}{\partial \phi^2} \\
    \frac{\partial^2 v'_m}{\partial \epsilon_{\theta}^2} = \frac{\partial^2 v'_m}{\partial \phi^2}
\end{gather*}
%

% Can use something like this to put references on a page
% by themselves when using endfloat and the captionsoff option.
% \ifCLASSOPTIONcaptionsoff
%   \newpage
% \fi

% trigger a \newpage just before the given reference
% number - used to balance the columns on the last page
% adjust value as needed - may need to be readjusted if
% the document is modified later
%\IEEEtriggeratref{8}
% The "triggered" command can be changed if desired:
%\IEEEtriggercmd{\enlargethispage{-5in}}

% references section

% can use a bibliography generated by BibTeX as a .bbl file
% BibTeX documentation can be easily obtained at:
% http://mirror.ctan.org/biblio/bibtex/contrib/doc/
% The IEEEtran BibTeX style support page is at:
% http://www.michaelshell.org/tex/ieeetran/bibtex/
%\bibliographystyle{IEEEtran}
% argument is your BibTeX string definitions and bibliography database(s)
%\bibliography{IEEEabrv,../bib/paper}
%
% <OR> manually copy in the resultant .bbl file
% set second argument of \begin to the number of references
% (used to reserve space for the reference number labels box)

\bibliography{IEEEabrv.bib,bibliography.bib}{}
\bibliographystyle{IEEEtran}

% \printbibliography

%%%%%%%%%%%%%%%%%%%%%%%%%%%%%%%%%%%%%
% References 
% \bibliographystyle{IEEEtran}
% \bibliography{IEEEabrv,./bibliography.bib}

%%%%%%%%%%%%%%%%%%%%%%%%%%%%%%%%%%%%%

% biography section
% 
% If you have an EPS/PDF photo (graphicx package needed) extra braces are
% needed around the contents of the optional argument to biography to prevent
% the LaTeX parser from getting confused when it sees the complicated
% \includegraphics command within an optional argument. (You could create
% your own custom macro containing the \includegraphics command to make things
% simpler here.)
%\begin{IEEEbiography}[{\includegraphics[width=1in,height=1.25in,clip,keepaspectratio]{mshell}}]{Michael Shell}
% or if you just want to reserve a space for a photo:

\begin{IEEEbiography}{Martijn Heller}
Biography text here.
\end{IEEEbiography}

% if you will not have a photo at all:
\begin{IEEEbiographynophoto}{Nikita Petrov}
Biography text here.
\end{IEEEbiographynophoto}

% insert where needed to balance the two columns on the last page with
% biographies
%\newpage

\begin{IEEEbiographynophoto}{Alexander Yarovoy}
Biography text here.
\end{IEEEbiographynophoto}

% You can push biographies down or up by placing
% a \vfill before or after them. The appropriate
% use of \vfill depends on what kind of text is
% on the last page and whether or not the columns
% are being equalized.

%\vfill

% Can be used to pull up biographies so that the bottom of the last one
% is flush with the other column.
%\enlargethispage{-5in}

% that's all folks
\end{document}